\pdfoutput=1
 \documentclass[conference]{IEEEtran}
\IEEEoverridecommandlockouts
\usepackage{cite}
\usepackage{amsthm}
\usepackage{amsmath,amssymb,amsfonts}
\usepackage{algorithmic}
\usepackage{graphicx}
\usepackage{textcomp}
\usepackage{xcolor}
\usepackage[noend,ruled,linesnumbered]{algorithm2e}
\usepackage{verbatim}
\usepackage{subfigure}
\usepackage{url}
\usepackage{enumitem}
\usepackage{xspace}
\usepackage{verbatim}
\usepackage{etoolbox}
\usepackage{hyperref}
\usepackage{pdfpages}
\makeatletter
\def\UrlAlphabet{%
      \do\a\do\b\do\c\do\d\do\e\do\f\do\g\do\h\do\i\do\j%
      \do\k\do\l\do\m\do\n\do\o\do\p\do\q\do\r\do\s\do\t%
      \do\u\do\v\do\w\do\x\do\y\do\z\do\A\do\B\do\C\do\D%
      \do\E\do\F\do\G\do\H\do\I\do\J\do\K\do\L\do\M\do\N%
      \do\O\do\P\do\Q\do\R\do\S\do\T\do\U\do\V\do\W\do\X%
      \do\Y\do\Z}
\def\UrlDigits{\do\1\do\2\do\3\do\4\do\5\do\6\do\7\do\8\do\9\do\0}
\g@addto@macro{\UrlBreaks}{\UrlOrds}
\g@addto@macro{\UrlBreaks}{\UrlAlphabet}
\g@addto@macro{\UrlBreaks}{\UrlDigits}
\makeatother

\def\BibTeX{{\rm B\kern-.05em{\sc i\kern-.025em b}\kern-.08em
    T\kern-.1667em\lower.7ex\hbox{E}\kern-.125emX}}
    
\BeforeBeginEnvironment{figure}{\vskip-0.5ex}
\AfterEndEnvironment{figure}{\vskip-0.5ex}
\theoremstyle{definition}
\newtheorem{theorem}{Theorem}
\newtheorem{example}{Example}
\newtheorem{definition}{Definition}
\newtheorem*{remark}{Remark}
\newtheorem*{soundness}{Soundness}

\newcommand{\stitle}[1]{\vspace{1ex} \noindent{{\bf #1}}}

\newcommand{\kw}[1]{{\ensuremath {\mathsf{#1}}}\xspace}

\newcommand{\cfl}{\kw{CFL}}
\newcommand{\daf}{\kw{DAF}}
\newcommand{\ceci}{\kw{CECI}}
\newcommand{\gpsm}{\kw{GpSM}}
\newcommand{\gsi}{\kw{GSI}}
\newcommand{\fast}{\kw{FAST}}
\newcommand{\cst}{\kw{CST}}
\newcommand{\best}{\kw{BEST}}
\newcommand{\average}{\kw{AVG}}
\newcommand{\worst}{\kw{WORST}}
\newcommand{\controlspace}{\vspace{-0.5\baselineskip}}

\begin{document}

\title{FAST: FPGA-based Subgraph Matching on Massive Graphs}

\author{{Xin Jin$^{\dagger}$, Zhengyi Yang$^{\S}$, Xuemin Lin$^{\S}$}, Shiyu Yang$^{\natural}$, Lu Qin$^{\ddagger}$, You Peng$^{\S}$%
\vspace{1.6mm}\\
\fontsize{10}{10}
\selectfont\itshape
$^\dagger$East China Normal University, $^\S$University Of New South Wales, $^\natural$Guangzhou University, $^\ddagger$University of Technology Sydney\\
\fontsize{9}{9} \selectfont\ttfamily\upshape
xinjin@stu.ecnu.edu.cn, \{zyang,lxue\}@cse.unsw.edu.au,\\ syyang@gzhu.edu.cn, lu.qin@uts.edu.au, you.peng@unsw.edu.au}

\maketitle
\begin{abstract}
Subgraph matching is a basic operation widely used in many applications.
However, due to its NP-hardness and the explosive growth of graph data, it is challenging to compute subgraph matching, especially in large graphs. In this paper, we aim at scaling up subgraph matching on a single machine using FPGAs. 
Specifically, we propose a CPU-FPGA co-designed framework. 
On the CPU side, we first develop a novel auxiliary data structure called \underline{c}andidate \underline{s}earch \underline{t}ree (\cst) which serves as a complete search space of subgraph matching. \cst can be partitioned and fully loaded into FPGAs' on-chip memory. Then, a workload estimation technique is proposed to balance the load between the CPU and FPGA.
On the FPGA side, we design and implement the first \underline{F}PGA-b\underline{a}sed \underline{s}ubgraph ma\underline{t}ching algorithm, called \fast. To take full advantage of the pipeline mechanism on FPGAs, task parallelism optimization and task generator separation strategy are proposed for \fast, achieving massive parallelism. Moreover, we carefully develop a BRAM-only matching process to fully utilize FPGA's on-chip memory, which avoids the expensive intermediate data transfer between FPGA's BRAM and DRAM.
Comprehensive experiments show that \fast achieves up to 462.0x and 150.0x speedup compared with the state-of-the-art algorithm \emph{DAF} and \emph{CECI}, respectively. In addition, \fast is the only algorithm that can handle the billion-scale graph using one machine in our experiments.
\end{abstract}

\section{Introduction}
Graph analysis has been playing an increasingly important role in the area of data analytics in recent years. One of the most fundamental problems in graph analysis is subgraph matching. 
Given a query graph $q$ and a data graph $G$, it aims to find all subgraphs of $G$ that are isomorphic to $q$. It has a wide range of applications including protein-protein interaction networks analysis \cite{prvzulj2006efficient}, chemical sub-compound search \cite{yan2004graph}, social network analysis \cite{snijders2006new}, computer aided design \cite{ohlrich1993subgemini}, and graph pattern mining \cite{teixeira2015arabesque}. It is also a core operation in graph databases \cite{neo4j} and RDF engines\cite{gstore}. However, it is challenging to compute subgraph matching, especially in large graphs, due to its NP-hardness \cite{hartmanis1982computers}. 

Extensive research has been conducted to develop efficient solutions for subgraph matching. Most practical solutions on CPUs \cite{cordella2004sub,shang2008taming,he2008graphs,zhao2010graph,han2013turboiso,bi2016efficient,bhattarai2019ceci,han2019efficient} 
are based on the backtracking approach, which recursively extends a partial embedding by mapping the next query vertex to a data vertex. Limited by the stand-alone design, these sequential solutions show unsatisfactory response time and poor scalability when handling massive graphs. 
%
In addition, general-purpose CPUs are not an ideal way to handle graph processing: they do not offer flexible high-degree parallelism, and their caches do not work effectively for irregular graphs with limited data locality.

\stitle{FPGAs.} 
%
FPGAs, which provide a new alternative to accelerate computation in the hardware level, has evolved rapidly in recent years. FPGAs have shown enormous advantages over CPUs on parallelism. Data can be directly streamed to FPGAs without instruction decoding and processed in pipelines. Because of its high potential to express parallelism at a massive scale and other benefits such as more energy-efficient than GPUs \cite{besta2019graph}, FPGAs have been 
applied to implement complex systems in industry. For example, Microsoft used FPGAs to speed up Bing Search and Azure Machine Learning \cite{microsoftcatapult}. FPGAs have also been rolled out by major cloud service providers such as Amazon Web Services \cite{amazon}, Alibaba \cite{alibaba}, Tencent \cite{tencent}, Huawei \cite{huawei}, and Nimbix \cite{nimbix}.  In academia, it has become a promising trend to use FPGAs to speed up different research problems including many graph processing problems \cite{besta2019substream,zhou2017accelerating,nurvitadhi2014graphgen,engelhardt2016gravf,zhou2018fpga}. 
Nevertheless, subgraph matching algorithms using FPGAs have not been developed in the literature.
FPGA-based subgraph matching can speed up and benefit all aforementioned applications. It can also be integrated into existing graph database systems (e.g. Neo4j \cite{neo4j}) and RDF engines (e.g. gStore\cite{gstore}) to accelerate various subgraph queries.

Motivated by this, in this paper, we explored how the pipeline mechanism of FPGAs can be fully utilized to accelerate the subgraph matching problem.

\stitle{Challenges.} We present the challenges of solving the problem of subgraph matching on FPGAs as follows:

\begin{itemize}[leftmargin=*]
\item \textit{Strictly pipelined design on FPGA}. FPGAs utilize a pipelined design, in which a fully pipelined loop demands no data dependencies among iterations. Thus the existing backtracking-based algorithms cannot be directly implemented on FPGAs. Furthermore, as FPGAs have an order of lower clock frequency than CPUs (e.g., 300MHz vs. 2GHz), it requires intricate design of the subgraph matching units on FPGAs to obtain high performance.

\item \textit{Limited FPGA on-chip memory}. FPGAs have small sizes of on-chip memory (BRAM) that are usually only tens of megabytes; hence the huge graph data and intermediate results will easily overflow BRAM when performing subgraph matching on FPGAs. Moreover, as fetching data from FPGA's external memory (DRAM) takes much more cycles than BRAM (e.g., 8 cycles vs. 1 cycle), frequent data transfer between BRAM and DRAM can significantly harm the performance. Thus, it is rather challenging to manage the data on FPGAs efficiently such that we can reduce the data transfer operations between BRAM and DRAM.
\end{itemize}

\stitle{Contributions.} To address these challenges, we propose a CPU-FPGA co-designed architecture which accelerates subgraph matching on a single machine using the power of FPGAs. Specifically, our main contributions are as follows.

\begin{itemize}[leftmargin=*]
\item \textit{The first CPU-FPGA co-designed framework to accelerate subgraph matching.} The framework includes a well-designed scheduler on the host side (i.e., the CPU) and a fully pipelined matching algorithm \fast on the kernel side (i.e., the FPGA). A workload estimation method is proposed on the host side for load-balancing between the CPU and FPGA, which can be exploited to extend our framework to multi-FPGA environment. To further improve the efficiency of \fast algorithm on the kernel side, we propose two optimizations with task parallelism and task generator separation.

\item \textit{A BRAM-only matching process to fully utilize FPGA's on-chip memory.} We first design an auxiliary data structure \cst to serve as a complete search space. An efficient partition strategy of \cst is proposed so that \cst can be fully loaded into BRAM, reducing the costly data fetching from FPGA's external memory. Then we propose a BRAM-only partial results buffer to avoid the expensive intermediate data transfer between BRAM and DRAM.

\item \textit{Extensive experiments using the industrial-standard LDBC benchmark.} Our experiments using LDBC \cite{ldbcbenchmark} show that \fast outperforms the state-of-the-art algorithms by orders of magnitude (up to 150.0x and 462.0x compared with \textit{CECI} \cite{bhattarai2019ceci} and \textit{DAF} \cite{han2019efficient}, respectively). More importantly, \fast is the only algorithm that can scale to the billion-scale graph on a single machine in our experiment.
\end{itemize}

\stitle{Paper Organization.} The rest of the paper is organized as follows. Section \ref{background} introduces background and Section \ref{related_work} presents related works. The system overview of the proposed solution is introduced in Section \ref{system_overview}, followed by the detailed design of software and hardware in Section \ref{software_implementation} and Section \ref{hardware_implementation}, respectively. Experimental results are presented in Section \ref{experiments}. Section \ref{conclusion} concludes the paper.
\section{Background}
\label{background}
In this section, the problem definition of subgraph matching is stated first, followed by a brief introduction of FPGAs. 

\subsection{Problem Definition}
A graph $G$ is represented as a tuple $G=(V,E,l,\Sigma)$, where $V(G)$ is the set of vertices, $E(G)\subset V\times V$ is the set of edges in G, $\Sigma$ is the set of labels, and $l$ is a labelling function that assigns each vertex $v\in V$ a label in $\Sigma$, denoted $l_G(v)$. We focus on \emph{undirected}, \emph{labelled}, \emph{connected}, and \emph{simple} graphs in this paper. 
Note that, our techniques can be readily extended to edge-labeled and directed graphs. We denote the number of vertices and edges in $G$ by $|V(G)|$ and $|E(G)|$, respectively. The set of neighbors of $v\in V(G)$ in $G$ is denoted by $N_G(v)=\{v'\in V(G)|(v, v')\in E(G)\}$ and the degree of $v$, denoted by $d_G(v)$, that is $d_G(v)=|N_G(v)|$. The $\overline{d}_{G}=\frac{2|E(G)|}{|V(G)|}$ and $D_G$ are denoted as the average and maximum degree, respectively.

\begin{definition}
(Subgraph Isomorphism) 
Given a query graph $q$ and a data graph $G$, 
$q$ is subgraph isomorphism to $G$ if and only if there is an \textit{injective} mapping $M$ from $V(q)$ to $V(G)$ such that $\forall u\in V(q)$, $l_q(u)=l_G(M(u))$ and $\forall (u, u')\in E(q)$, $(M(u), M(u'))\in E(G)$, where $M(u)$ is the vertex to which $u$ is mapped. 
\end{definition}

We refer to each injective mapping $M$ as a \emph{subgraph isomorphism embedding} of $q$ in $G$. A graph $g'$ is an \emph{induced subgraph} of $g$ if and only if $\forall \mu,\mu' \in V_{g'}, e=(\mu,\mu')\in E_g$, we have $e\in E_{g'}$. We call an embedding of an induced subgraph of $q$ in $G$ a \emph{partial embedding}, denoted as $p$. The $M_p(u)$ denotes the mapping vertex of $u$ in $q$. We use $\mathcal{O}$ to denote the \emph{matching order}, which is a sequence of query vertices representing the order they are matched.

\begin{figure}[htbp]
\controlspace
\centering
\subfigure[Query graph]{ \label{fig_query}
\includegraphics[scale=0.38]{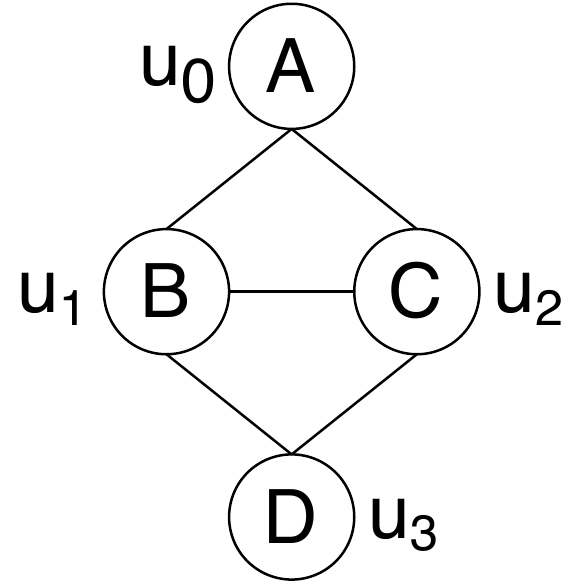}}
\hspace{0.3cm}
\subfigure[Data graph]{\label{fig_data}\includegraphics[scale=0.38]{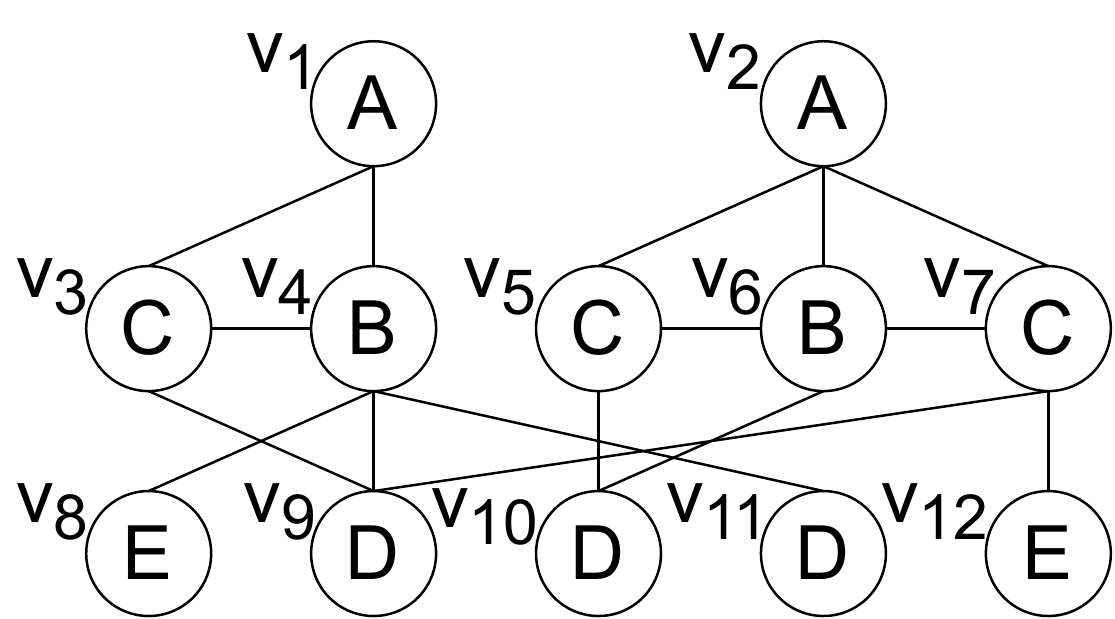}}
\controlspace
\caption{Subgraph Matching}
\label{fig_graph}
\controlspace
\controlspace
\end{figure}

\begin{example}
For example, consider the query graph $q$ in Fig. \ref{fig_query} and the data graph $G$ in Fig. \ref{fig_data}. Suppose the matching order is $\{u_0,u_1,u_2,u_3\}$, since there is a subgraph isomorphism embedding $M=\{(u_0,v_1),(u_1,v_4),(u_2,v_3),(u_3,v_9)\}$, $q$ is subgraph isomorphism to $G$. We call $p=\{(u_0,v_1),(u_1,v_4),(u_2,v_3)\}$ a partial embedding and $u_3$ will be the next query vertex to match.
\end{example}

\stitle{Problem Statement.} Given a query graph $q$ and a data graph $G$, we study the problem of subgraph matching, which efficiently extracts all subgraph isomorphic embeddings of $q$ in $G$.

\subsection{Characteristics of FPGA}
A field-programmable gate array (FPGA) is an integrated circuits that consists of a matrix of configurable logic, memory, and digital signal processing (DSP) components. These components are distributed within a grid of configurable routing wires connected to programmable chip I/O blocks. This flexible and programmable fabric can be configured to perform any functionality implemented as a digital circuit. The FPGA program statements are translated into a netlist of primitive components first and then be assigned to physical components in the FPGA fabric, determining which routing wires should be used to connect them. This architecture allows data to be directly streamed to FPGAs with no need to decode instructions, as necessary in CPUs, to achieve high efficiency.

FPGAs have an unique programming model in which computations are laid out spatially and the programmer has to specify how data and control flows from one logic block to another inside the data path. 
Thus, common design challenges when developing FPGA-based algorithms is the amount of space (resources) required and the ability to meet timing (ensuring the data can be moved across the circuit in a correct manner). It is also worth to mention that the clock rate on FPGAs is usually about 10x slower than that of CPUs (e.g., 300MHz vs 2GHz). 
Thus, FPGA-based algorithms must be thoughtfully designed to provide better performance than CPU implementations, by exploiting massive parallelism, typically in the form of deep pipelines. 

\section{Related Work}
\label{related_work}
\subsection{Subgraph Matching}

\stitle{Stand-alone Solutions.} 
The study of practical subgraph matching algorithms was initiated by Ullmann's backtracking algorithm \cite{ullmann1976algorithm}, which recursively matches query vertices to data vertices following a given matching order. 
Later researches \cite{cordella2004sub,shang2008taming,he2008graphs,zhao2010graph} focus on different matching order, pruning rules, and index structure.
Turbo$_{ISO}$ \cite{han2013turboiso} proposes to merge similar vertices with and a \emph{CR} index structure. 
CFL-Match \cite{bi2016efficient} proposes the core-forest-leaf decomposition to reduce redundant Cartesian products and proposes a more compact auxiliary structure \emph{CPI} to solve the exponential size of \emph{CR}.
CECI \cite{bhattarai2019ceci} and DAF \cite{han2019efficient} adopt the intersection-based method to find the candidates, which demonstrate better performance than the edge verification method used in previous works \cite{sun2020memory}. However, these solutions fail to accommodate large graphs due to their inherent sequential nature.

\stitle{Distributed Solutions.} 
Most distributed algorithms utilize distributed join to compute matches \cite{lai2019distributed}. \cite{lai2015scalable,lai2016scalable,qiao2017subgraph} decompose the query graph into sub-queries, find the matches of each sub-query, and use a series of binary joins to assemble the final results. \cite{ammar2018distributed}, on the other hand, grows the query graph one vertex at a time following a specific order to obtain worst-case optimality.
\fast can be potentially used to accelerate the computation in distributed subgraph matching.

\stitle{GPU-based Solutions.} 
GpSM \cite{tran2015fast} and GunrockSM \cite{wang2016fast} adopt the binary join strategy in GPUs, which collects candidates for each edge of $q$ and joining them to find final matches. They suffer from high computation workload, high memory latency, and severe workload imbalance. GSI \cite{zeng2020gsi} proposes a Prealloc-Combine approach, which joins candidate vertices instead of edges to improve the efficiency. The algorithms mentioned above are only able to handle the graphs that can be fit into the GPU memory. PBE \cite{guo2020gpu} solves this by partitioning the graph in advance and matching intra- and inter-partition matches in two separate steps. However, as the on-chip memory of an FPGA is order-of-magnitude smaller than a GPU memory, this approach can hardly be applied to FPGAs.

\subsection{FPGA-based Acceleration of Graph Processing}
FPGAs can be an energy-efficient solution to deliver specialized hardware for graph processing. This is reflected by the recent interests in developing various graph algorithms and graph processing frameworks on FPGAs. For examples, \cite{besta2019substream} applies FPGAs to speed up \emph{Maximum Matching} and \cite{zhou2017accelerating} utilizes FPGAs to accelerate the process of the \emph{Single-Source-Shortest-Paths}. In addition to these specific graph algorithms on FPGAs, a lot of effort was devoted to design generic frameworks for facilitating the implementation of graph algorithms on FPGAs \cite{nurvitadhi2014graphgen,engelhardt2016gravf,zhou2018fpga}. However, these frameworks are usually built upon specific programming models (e.g. BSP, Vertex-Centric) supporting only limited APIs. This restricts the implementation of a highly optimized subgraph matching algorithm. More critically, most of the frameworks can only handle small graphs and cannot scale to large ones.


\section{System Overview}
\label{system_overview}
\begin{figure}[htbp]
\controlspace
\centerline{\includegraphics[width=0.8\columnwidth]{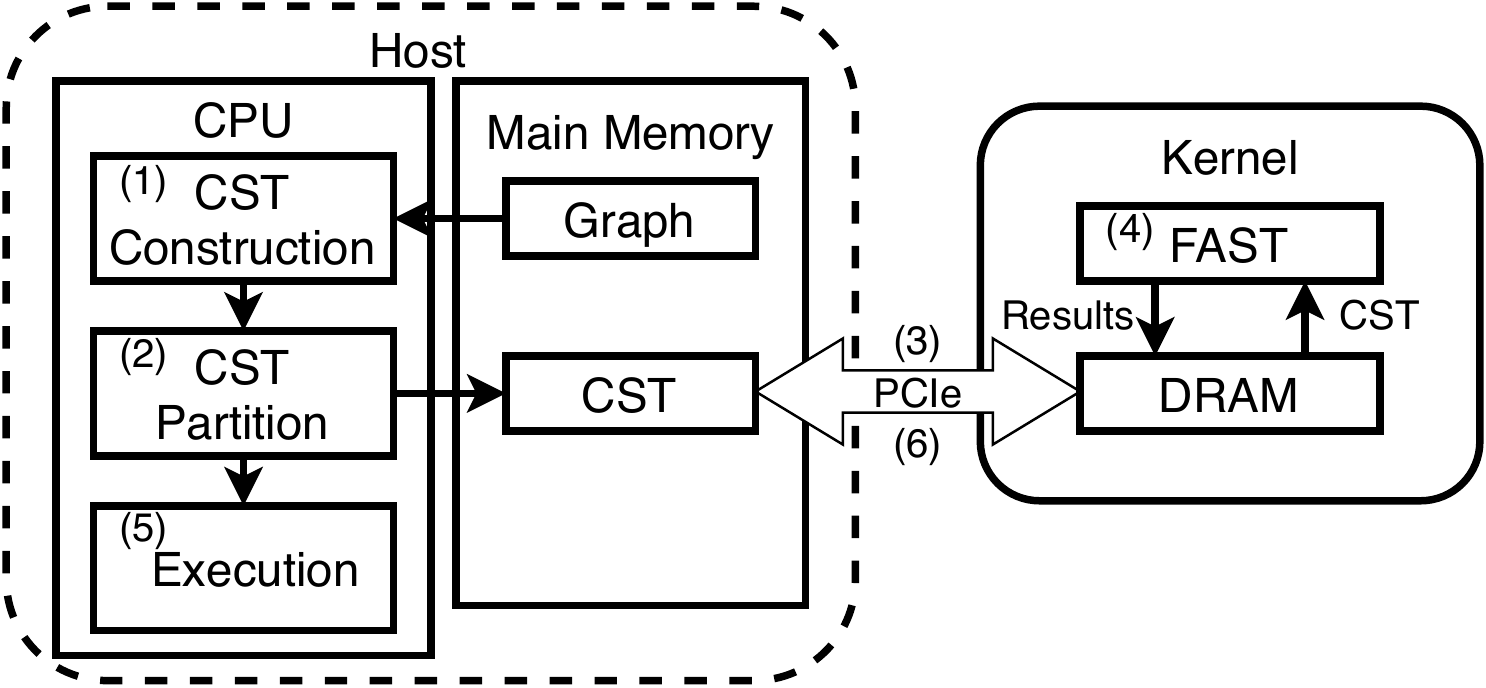}}
\controlspace
\caption{The overall system architecture}
\label{architucture}
\controlspace
\end{figure}

The overview architecture of our system is illustrated in Fig. \ref{architucture}. The host side, i.e. CPU, takes charge of constructing and partitioning our novel auxiliary data structure \cst and offloading them to FPGA through PCIe bus. It also shares a small portion of matching tasks to improve throughput. The kernel side, i.e. FPGA card\footnote{In this paper, we focus on FPGAs with DRAM attached, while our techniques can be applied on FPGAs without DRAM as well.}, is PCIe-attached to the host machine, focusing on the subgraph matching tasks.

When the query and data graph are read into the host's main memory, the system launches the execution tasks described as follows:

\begin{enumerate}[leftmargin=*]
\item CPU constructs \cst based on $q$ and $G$, which prunes a large number of false positives according to graph attributes such as labels and degrees, etc. \cst serves as a complete search space for all embeddings of $q$ in $G$ (Section \ref{software_cst}).
\item Limited by FPGA on-chip resources, \cst is often too large to be fully loaded into BRAM. The host side partitions \cst to satisfy the size constraint (Section \ref{software_partition}).
\item Once a partitioned \cst satisfies the constraint, it is transferred to DRAM on FPGA card from the host's main memory through PCIe bus.
\item On the kernel side, \fast reads a partitioned \cst from DRAM to BRAM and runs subgraph matching on it. The results are flushed to DRAM when the whole search space of this \cst has been searched. \fast repeats this procedure as long as there exists an unprocessed \cst (Section \ref{hardware_implementation}).
\item On the host side, when all \cst has been partitioned and offloaded, CPU shares a small portion of matching tasks to improve the overall throughput (Section \ref{software_share}).
\item When FPGA finishes its processing, CPU receives a termination signal and fetches results to the main memory.
\end{enumerate}
The details of software and hardware implementation are described in Section \ref{software_implementation} and Section \ref{hardware_implementation}, respectively.
\section{Software Implementation}
\label{software_implementation}
In this section, we first introduce our novel auxiliary data structure \cst and its partition strategy. Then we present how to schedule matching tasks between the host and kernel side.

\begin{table}[htbp]
\controlspace
\caption{definition of parameters in software implementation}
\begin{center}
\setlength{\tabcolsep}{7mm}
\begin{tabular}{|c|c|}
\hline
\textbf{Symbol}&\textbf{Definition} \\
\hline
\cst & candidate search tree \\
\hline
$t_q$ & a breadth-first search tree of $q$\\
\hline
$C(u)$ & the candidate set of $u$ in \cst \\
\hline
$N^u_{u'}(v)$ & the adjacency list of $v$ regarding $(u,u')$\\
\hline
$u_p$ / $u_c$ & the parent / child vertex of $u$ in \cst \\
\hline
$u_n$ & the non-tree neighbor of $u$ in \cst \\
\hline
$\mathcal{O}$ & the matching order of $q$ \\
\hline
$|\cst|$ & the size of \cst \\
\hline
$D_{\cst}$ &the maximum degree of candidates in \cst\\
\hline
\end{tabular}
\label{notations_in_sf}
\end{center}
\controlspace
\controlspace
\controlspace
\end{table}

\subsection{\cst Structure}
\label{software_cst}
We adopt the indexing-enumeration framework; that is, construct an auxiliary data structure, then compute all embeddings based on this data structure. 
Following conventional technique \cite{bi2016efficient,serafini2017qfrag}, the query graph is firstly transformed into a spanning tree.
Given a query graph $q$ and a data graph $G$, we build an auxiliary data structure upon them called \emph{candidate search tree} (\cst).

\begin{definition} (Candidate Search Tree) 
Given a query graph $q$ and a data graph $G$, a candidate search tree $\cst_{(q,G)}$ is a graph\footnote{We abuse the term \textit{tree} during naming to emphasize that \cst is constructed based on the spanning tree of the query.}
that is isomorphic to $q$. Each vertex $u$ of $\cst_{(q,G)}$ has a candidate set, denoted $C(u)$, which stores all vertices of $G$ that $u$ can be mapped. There is an edge between $v\in C(u)$ and $v'\in C(u')$ for adjacent vertices $u$ and $u'$ in $\cst_{(q,G)}$ if and only if $(v,v') \in E(G)$.
\end{definition}

We denote $\cst_{(q,G)}$ as \cst if the context is clear.
 Given the query graph $q$ and its BFS trees $t_q$, we call adjacent vertices $u$ and $u_n$ in \cst \emph{non-tree neighbors} if $(u,u_n)\in E(q)$ but $(u,u_n)\notin E(t_q)$. The adjacent candidates $v\in C(u)$ and $v_n\in C(u_n)$ for non-tree neighbors $u$ and $u_n$ in \cst are called \emph{non-tree candidate neighbors}. We use $N_{u'}^{u}(v)$ to denote the adjacency list of $v\in C(u)$ regarding $(u,u')$ in \cst, i.e., $N_{u'}^{u}(v)=\{v'\in C(u')\ |\ (v,v')\in E(\cst)\}$. \cst inherits the parent-child relationships of $t_q$. We use $u_p$ and $u_c$ to denote the parent and child vertex of $u$, respectively. The vertex $u$ in \cst is a \emph{leaf} or \emph{root} vertex if $u$ has no child vertices or parent vertices, respectively.

\begin{figure}[htbp]
\controlspace
\centering
\subfigure[BFS Tree $t_q$]{\label{fig_bfs_example}
\includegraphics[scale=0.38]{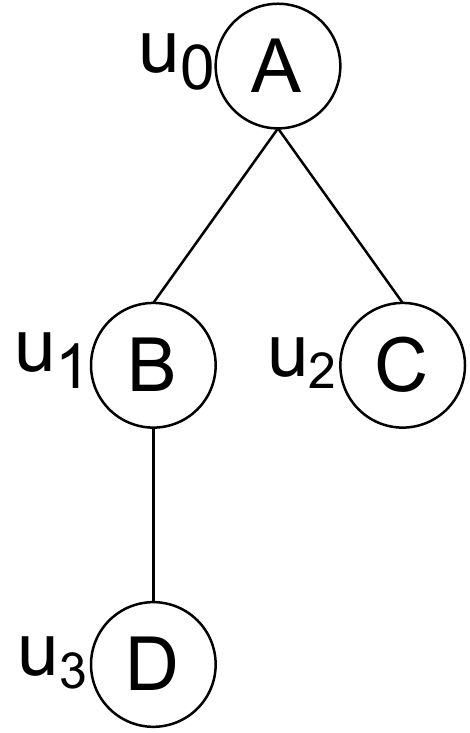}}
\hspace{1cm}
\subfigure[CST]{\label{fig_cst_example}\includegraphics[scale=0.38]{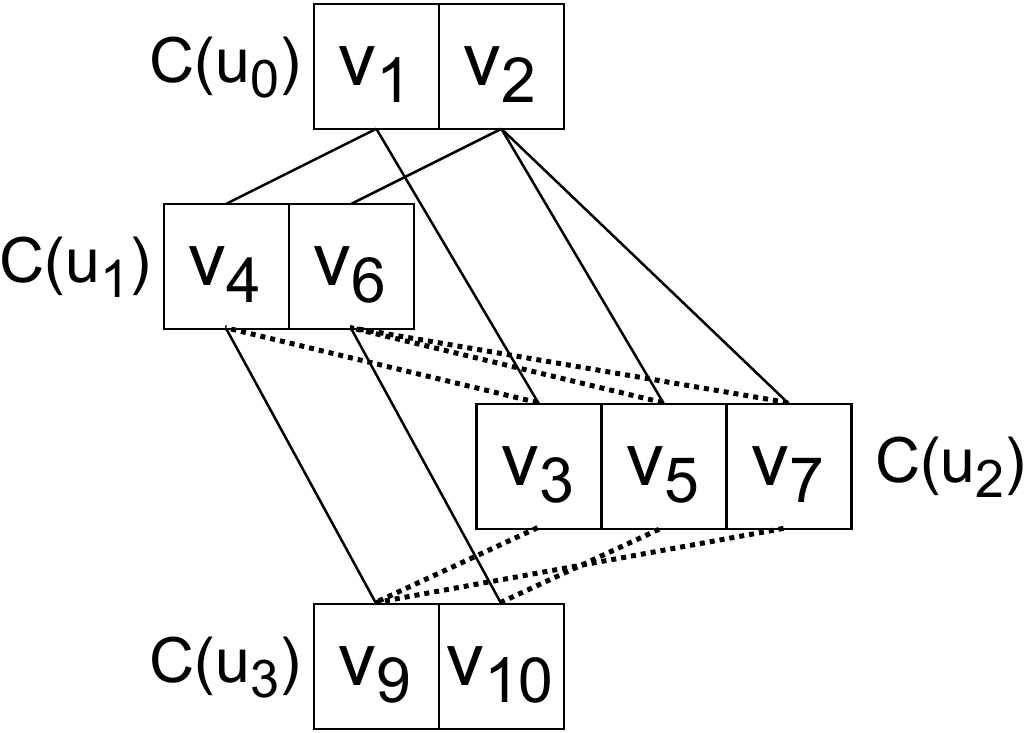}}
\controlspace
\caption{Example \cst structure}
\controlspace
\end{figure}

\begin{example}
For example, given the query graph $q$, the data graph $G$ in Fig. \ref{fig_graph} and BFS tree $t_q$ of $q$ in Fig. \ref{fig_bfs_example}, the corresponding \cst is in Fig. \ref{fig_cst_example}. Then $u_1$ and $u_2$ are called non-tree neighbors because $(u_1,u_2)\notin E(t_q)$, while $v_4\in C(u_1)$ and $v_3\in C(u_2)$ are called non-tree candidate neighbors. $C(u_1)=\{v_4,v_6\}$, $C(u_2)=\{v_3,v_5,v_7\}$, $N^{u_1}_{u_2}(v_6)=\{v_5,v_7\}$ and $N^{u_2}_{u_3}(v_3)=\{v_9\}$. The all embeddings of $q$ in $G$ $\{(u_0,v_1),(u_1,v_4),(u_2,v_3),(u_3,v_9)\}$ and $\{(u_0,v_2),(u_1,v_6),(u_2,v_5),(u_3,v_{10})\}$ can be computed by traversing only the \cst.
\end{example}

\controlspace
\begin{algorithm}[htb]\label{alg_cst_constructor}
\small
\caption{\cst{Constructor}($q,G,t_q$)}
\KwIn{$q,G,t_q$}
\KwOut{\cst}
$root\leftarrow$ root vertex of $t_q$\;
$C(root)\leftarrow$ compute candidates of $root$\;\label{alg_cst_verify_cand_1}
\tcc{Line \ref{alg_cst_top_start}-\ref{alg_cst_top_end}: Top-Down Construction}
\ForEach{$u\in V(q)$ in a top-down fashion}{\label{alg_cst_top_start}
    $C(u)\leftarrow$ compute candidates of $u$\;\label{alg_cst_verify_cand_2}
    \ForEach{$v_p\in C(u_p)$}{
        \ForEach{$v\in C(u)$}{
            \lIf{$(v,v_p)\in E(G)$}{$N_u^{u_p}(v_p).push(v)$}
        }
    }
}\label{alg_cst_top_end}
\BlankLine
\tcc{Line \ref{alg_cst_bottom_start}-\ref{alg_cst_bottom_end}: Bottom-Up Refinement}
\ForEach{$u\in V(q)$ in a bottom-up fashion}{\label{alg_cst_bottom_start}
    \ForEach{$v\in C(u)$}{
        \If{$v$ is not valid}{\label{alg_cst_valid_start}
            remove $v$ and its adjacency lists\;
        }\label{alg_cst_valid_end}
        \ForEach{child vertex $u_c$ of $u$ in $t_q$}{
            \ForEach{$v'\in N_{u_c}^{u}(v)$}{
                \lIf{$v'\notin C(u_c)$}{remove $v'$ from  $N_{u_c}^{u}(v)$}
            }
        }
    }
}\label{alg_cst_bottom_end}
\BlankLine
\tcc{Line \ref{alg_cst_non_start}-\ref{alg_cst_non_end}: Add Edges Between Non-tree Candidate Neighbors}
\ForEach{$u\in V(q)$}{\label{alg_cst_non_start}
    \ForEach{$v\in C(u)$}{
        \ForEach{non-tree neighbor $u_n$ of $u$}{
            \ForEach{$v_n\in C(u_n)$}{
                \lIf{$(v,v_n)\in E(G)$}{$N_{u_n}^{u}(v).push(v_n)$}
            }
        }
    }
}\label{alg_cst_non_end}
\Return{\cst}
\end{algorithm}
\controlspace
\controlspace

The construction of \cst is described in Alogrithm \ref{alg_cst_constructor}. We first adopt the similar top-down construction (Line \ref{alg_cst_top_start}-\ref{alg_cst_top_end}) and bottom-up refinement (Line \ref{alg_cst_bottom_start}-\ref{alg_cst_bottom_end}) in \cite{bi2016efficient} to build a tree-like data structure. We verify whether a data vertex conforms with the local features of the query vertex to compute candidate set $C(u)$ (Line \ref{alg_cst_verify_cand_1}, Line \ref{alg_cst_verify_cand_2}). A candidate $v$ of vertex $u$ is valid if $|N^u_{u_c}(v)|\neq 0$ for any child vertex $u_c$ of $u$ and $\exists v_p\in C(u_p)$ that $v\in N^{u_p}_{u}(v_p)$. We remove $v$ from $C(u)$ and its adjacency lists if $v$ is not valid during the bottom-up refinement. (Line \ref{alg_cst_valid_start}-\ref{alg_cst_valid_end}). Then edges are added between non-tree candidate neighbors (Line \ref{alg_cst_non_start}-\ref{alg_cst_non_end}). 

\begin{figure*}[htbp]
\centering
\subfigure[Initial \cst with $k=2$] {\label{fig_init_cst}\includegraphics[scale=0.35]{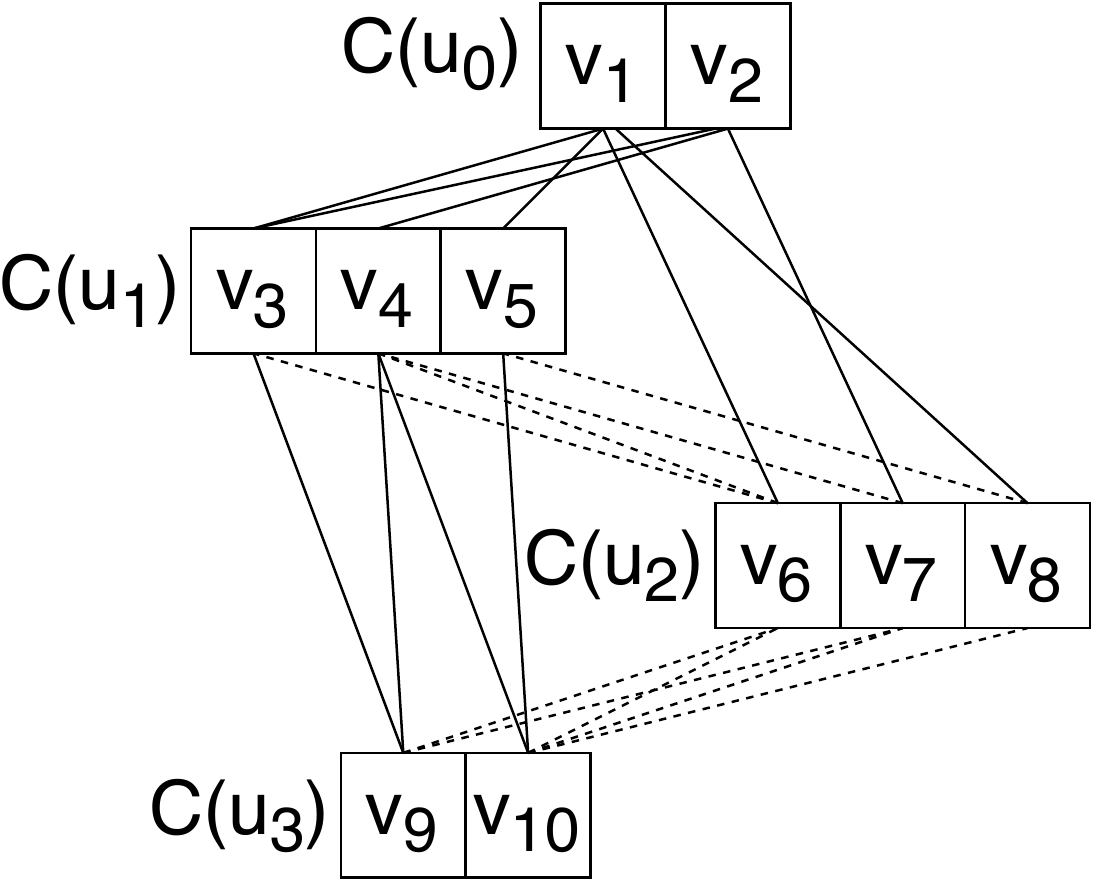}}
\hspace{0.1cm}
\subfigure[1st partition of \cst]{ \label{fig_partition_a}
\includegraphics[scale=0.35]{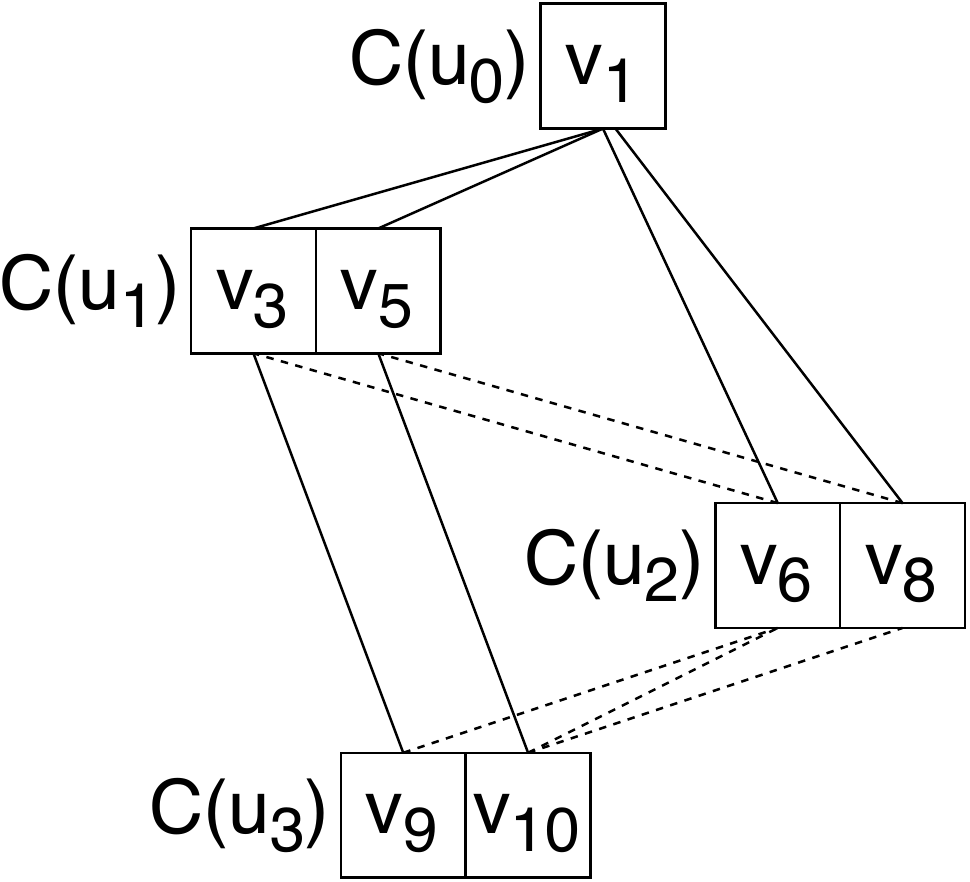}}
\hspace{0.1cm}
\subfigure[2nd partition of \cst]{\label{fig_partition_b}\includegraphics[scale=0.35]{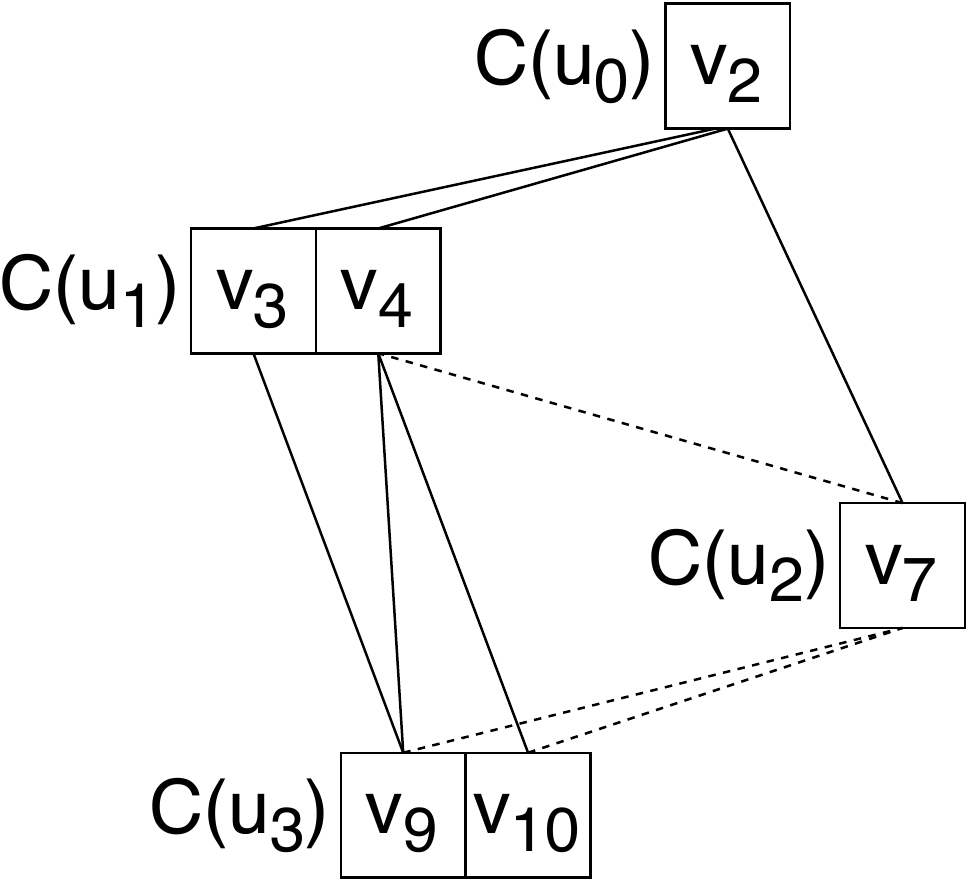}}
\hspace{0.1cm}
\subfigure[Workload Estimation] {\label{fig_workload}\includegraphics[scale=0.35]{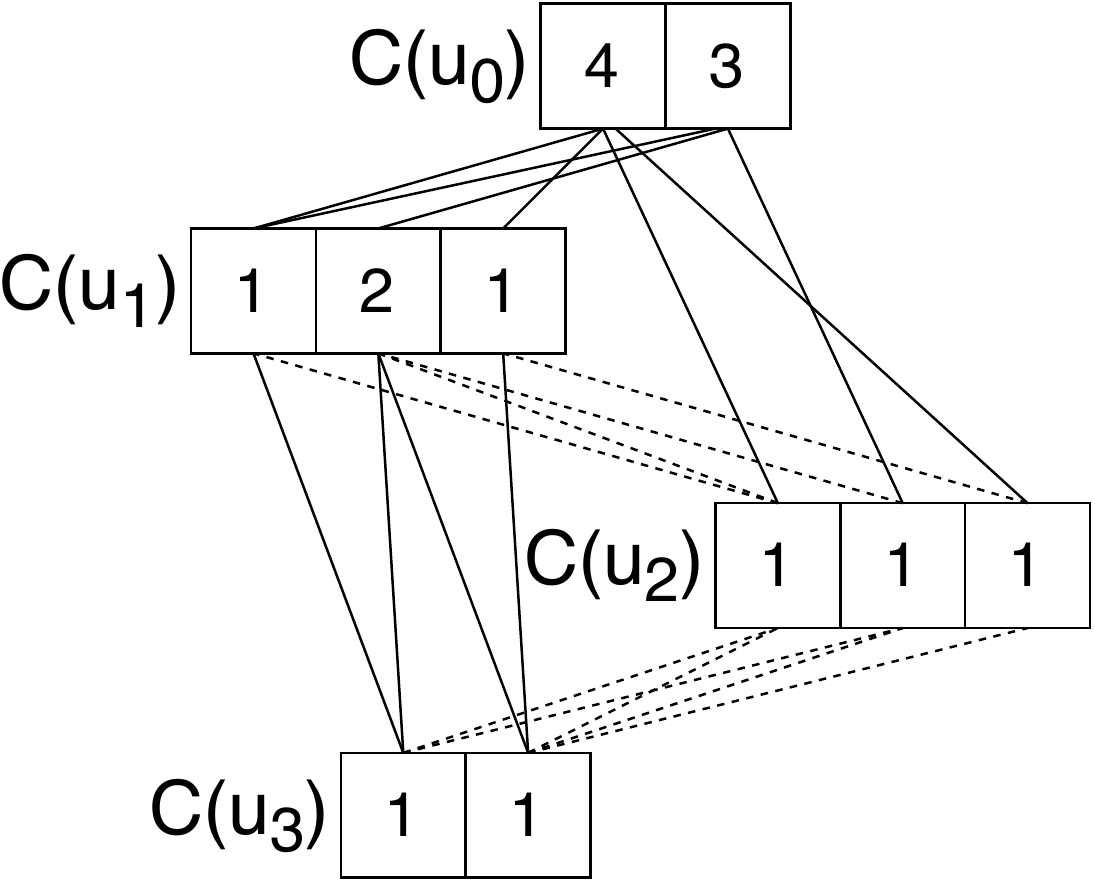}}
\vspace{-0.2cm}
\caption{Running Example of Scheduling}
\label{fig_example}
\vspace{-0.8\baselineskip}
\end{figure*}

\begin{soundness} 
\cst should serve as a complete search space for the given query graph $q$ over the data graph $G$. To achieve this, \cst must satisfy the following soundness constraint:
\begin{itemize}[leftmargin=*]
\item 
\emph{For every vertex $u$ in \cst, if there is an embedding of $q$ in $G$ that maps $u$ to $v$, then $v$ must be in $C(u)$.}
\end{itemize}
Note that, although in the soundness requirement we only consider candidates of query vertices, the edges between candidates are automatically included based on our \cst definition. Regarding a sound \cst, we have the following theorem.
\begin{theorem}
Given a sound \cst, all embeddings of $q$ in $G$ can be computed by traversing only the \cst. 
\end{theorem}
\end{soundness}

\begin{remark}
\cst has vital differences with the auxiliary data structure in previous works, namely \emph{CPI (compact path-index)} \cite{bi2016efficient} and \emph{CS (candidate space)} \cite{han2019efficient}. Compared with \emph{CPI}, \cst uses all edge information in $q$ during construction (by adding non-tree edges), making it a complete search space. Hence, it can be partitioned and the embeddings of each partition can be computed independently in FPGA's BRAM (details will be introduced in the next subsection). The reasons that we do not use the structure \emph{CS} are as follows: (1) The top-down construction and bottom-up refinement of \cst is equivalent to the first two refinements (totally three) of \emph{CS}, making the size of \cst close to \emph{CS} for most data graphs;
(2) Constructing \cst is much less expensive because the edges between non-tree candidate neighbors are not updated during construction as necessary in \emph{CS}. Consequently, \cst can potentially have a larger search space than \emph{CS} because of fewer pruning steps. However, there is an essential trade-off between the size of search space and the construction cost. Compared with pure CPU-based algorithms, \fast is more sensitive to the cost of constructing the auxiliary data structure conducted by CPU, to let FPGA receive its tasks from the host as soon as possible.
\end{remark}
\subsection{\cst Partition}
\label{software_partition}
Limited by on-chip resources on FPGAs, \cst is often too large to be fully loaded into BRAM. Generally, the read latency of BRAM is 1 cycle while DRAM is about 7-8 cycles. Our experiments show the dramatic performance decreasing when we access \cst from DRAM rather than BRAM (Section \ref{experiment_dram}). On the other hand, accesses to \cst are random and unpredictable, which eliminates the possibility of prefetching the data from DRAM to BRAM. Hence, it is necessary to partition \cst  and offload them to FPGA one by one.

In addition to the size of \cst, denoted as $|CST|$, we also set a limitation on the maximum degree of candidates in \cst, i.e., $D_{\cst}$. The reason is that the maximum number of access ports to an adjacency list are limited on FPGAs and it will be discussed in detail in Section \ref{hardware_implememtation_basic}. We use $\delta_S$ and $\delta_D$ to denote the threshold of $|\cst|$ and $D_{\cst}$, respectively. We partition the \cst if either $|CST|>\delta_S$ or $D_{CST}>\delta_D$.

The partition strategy of \cst is illustrated in Algorithm \ref{alg_cst_partition}. Note that we adopt the path-based method to compute the matching order $\mathcal{O}$ in this paper, which determines $\mathcal{O}$ by ordering the root-to-leaf paths of $t_q$. However, our method is designed to work with any arbitrary connected matching orders. Initially, to partition \cst, we partition candidates of root vertex in \cst. If there is only one candidate of root vertex in \cst, we move on to partition candidates of next vertex $u$ in $\mathcal{O}$. The first step is to determine the partition factor $k$, which equals to the maximum value between the ratio of $|CST|$ and $D_{CST}$ to their corresponding thresholds (Line \ref{alg_cst_partition_k_max}). If $k$ exceeds the number of candidates, i.e., $|C(u)|$, we set $k$ to $|C(u)|$ (Line \ref{alg_cst_partition_k_min}). We partition $C(u)$ into $k$ parts evenly and then construct a new \cst level-by-level in a top-down manner. For those vertices precedes $u$ in $\mathcal{O}$, we pick candidates as the same as the old \cst (Line \ref{alg_cst_partition_precede_start}-\ref{alg_cst_partition_precede_end}). For those vertices follows $u$ in $\mathcal{O}$, we pick candidates in old \cst which can reach at least one candidate in the partitioned $C(u)$ (Line \ref{alg_cst_partition_follow_start}-\ref{alg_cst_partition_follow_end}). \cst is offloaded to FPGA or assigned to CPU as soon as it satisfies $|CST|$ and $D_{\cst}$ constraints (Line \ref{alg_cst_partition_offload}). Otherwise, it will be further partitioned recursively. 

\begin{algorithm}[htbp]\label{alg_cst_partition}
\small
\caption{\cst{Partition}(\cst,$\mathcal{O},index$)}
\KwIn{\cst,$\mathcal{O},index$}
$u\leftarrow \mathcal{O}[index]$\;
$k\leftarrow max(\frac{|\cst|}{\delta_S},\frac{D_{\cst}}{\delta_D})$\;\label{alg_cst_partition_k_max}
$k\leftarrow min(k,\cst.|C(u)|)$\;\label{alg_cst_partition_k_min}
partition $\cst.C(u)$ into k parts evenly\;
\For{$i$ from $0$ to $k$}{
$\cst^{'}\leftarrow \emptyset$\;
\ForEach{vertex $u'$ precedes $u$ in $\mathcal{O}$}{\label{alg_cst_partition_precede_start}
$\cst^{'}.C(u')\leftarrow \cst.C(u')$\;
}\label{alg_cst_partition_precede_end}
\ForEach{vertex $u'$ follows $u$ in $\mathcal{O}$}{\label{alg_cst_partition_follow_start}
\ForEach{candidate $v$ in $\cst.C(u')$}{
\If{v can reach $i$th partitioned $\cst.C(u)$}{
$\cst^{'}.C(u')\leftarrow \cst^{'}.C(u')\cup \{v\}$\;
}
}
}\label{alg_cst_partition_follow_end}
update adjacency lists of $\cst^{'}$ based on \cst\;
\If{$|\cst^{'}|\leq\delta_{S}$ and $D_{\cst^{'}}\leq\delta_{D}$}{
\cst{Process}($\mathcal{O}$, $\cst^{'}$)}\;\label{alg_cst_partition_offload}

\ElseIf{$\cst^{'}.|C(u)|$ equals to $1$}{
CST{Partition}($\cst^{'},\mathcal{O},index+1$)\;
}\lElse{
CST{Partition}($\cst^{'},\mathcal{O},index$)}
}
\end{algorithm}

\begin{example}
As shown in Fig. \ref{fig_init_cst}, suppose $k$ is 2, we first partition the root candidates $\{v_1,v_2\}$ into 2 parts: $\{v_1\}$ and $\{v_2\}$. Then, to construct \cst rooted by $v_1$, we pick candidates of $u_1$ and $u_2$ that are adjacent to $v_1$, which are $\{v_3,v_5\}$ and $\{v_6,v_8\}$. After that, we pick the candidates of $u_3$ that can reach $v_1$, which are $\{v_9,v_{10}\}$. Obviously, there is no overlap of the search space between two partitioned \cst in Fig. \ref{fig_partition_a} and Fig. \ref{fig_partition_b}, so no repeated results will be reported. 
\end{example}

\subsection{Schedule the Matching Tasks}
\label{software_share}
After finishing the partition of \cst, the host side shares a small portion of matching tasks in order to further improve the throughput as a whole. Considering load balancing between the CPU and FPGA, the workload of \cst, denoted as $W_{\cst}$, should be estimated first. The size of search space in different \cst can usually differs a lot due to the power-law feature of real-world graphs. We use the number of embeddings in \cst without considering any false positives to estimate $W_{\cst}$. It can be computed in a bottom-up way using a dynamic programming algorithm. For each candidate $v\in C(u)$, we compute $c_u(v)$, the number of embeddings in \cst for the subgraph of $q$ induced by the suffix of matching order starting from $u$ such that $u$ is mapped to $v$. Initially, $c_u(v)=1$ for all leaf vertices $u$. Then we compute $c_u(v)$ in a bottom-up fashion, where $c_u(v)=\prod_{u'\in u.child}\sum_{v'\in N_{u'}^{u}(v)} c_{u'}(v')$. Finally, the total workload $W_{\cst}=\sum_{v\in C(u_r)}c_{u_r}(v)$.

\begin{example}
Given \cst in Fig. \ref{fig_init_cst} and $t_q$ in Fig. \ref{fig_bfs_example}, the workload estimation results are illustrated in Fig. \ref{fig_workload}. For leaf vertices $u_3$ and $u2$, $c_{u_3}(v_9)=c_{u_3}(v_{10})=c_{u_2}(v_6)=c_{u_2}(v_7)=c_{u_2}(v_8)=1$. Then we compute $c_u(v)$ in a bottom-up fashion, e.g., $c_{u_0}(v_1)=(c_{u_1}(v_3)+c_{u_1}(v_5))*(c_{u_2}(v_6)+c_{u_2}(v_8))=4$. Finally, $W_{\cst}=c_{u_0}(v_1)+c_{u_0}(v_2)=4+3=7$.
\end{example}

\controlspace
\controlspace
\begin{algorithm}[htb]\label{alg_cst_process}
\small
\caption{\cst{Process}($\mathcal{O},\cst$)}
\KwIn{$\mathcal{O},\cst$}
$W_{\cst}\leftarrow$compute the workload of \cst\;\label{alg_cst_process_workload}
\If{$W_C+W_{CST}<\delta\times (W_C+W_F+W_{\cst})$}
{
    assign \cst to CPU\;\label{alg_cst_process_if_start}
    $W_C\leftarrow W_C+W_{CST}$\;\label{alg_cst_process_if_end}
}\Else{
    FAST(\cst, $\mathcal{O}$)\;\label{alg_cst_process_else_start}
    $W_F\leftarrow W_F+W_{CST}$\;\label{alg_cst_process_else_end}
}
\end{algorithm}
\controlspace
\controlspace

As illustrated in Algorithm \ref{alg_cst_process}, we restrict the proportion of the total workload of \cst assigned to the host side from exceeding a threshold, denoted as $\delta$. When a valid \cst is constructed, $W_{\cst}$ is computed first (Line \ref{alg_cst_process_workload}). We use $W_C$ and $W_F$ to denote the total workload of \cst assigned to the host and kernel side, respectively. If $\frac{W_C+W_{\cst}}{W_C+W_F+W_{\cst}}<\delta$, it will be assigned to the host side (Line \ref{alg_cst_process_if_start}-\ref{alg_cst_process_if_end}). Otherwise, FPGA takes charge of this \cst (Line \ref{alg_cst_process_if_start}-\ref{alg_cst_process_if_end}). The host side uses the basic backtracking subgraph matching algorithm to process \cst. It should be noted that when \cst is assigned to CPU, \cst is temporarily cached and will be processed when all partition procedure finishes. When \cst is assigned to FPGA, \cst is offloaded to FPGA immediately.
\section{Hardware Implementation}
\label{hardware_implementation}
In this section, we first present our proposed algorithm \fast to accelerate subgraph matching on FPGAs. Then we introduce several important optimizations to improve the matching process based on FPGA characteristics. Notations of all the related parameters are listed in Table \ref{notations}.

\begin{table}[htbp]
\controlspace
\controlspace
\caption{definition of parameters}
\controlspace
\controlspace
\begin{center}
\setlength{\tabcolsep}{6.5mm}
\begin{tabular}{|c|c|}
\hline
\textbf{Symbol}&\textbf{Definition} \\
\hline
$\mathcal{P}$ & Intermediate results buffer \\
\hline
$\mathcal{M}$ & the set of all embeddings \\
\hline
$p_{i}$ / $p_{o}$ & An input / output partial result \\
\hline
$\mathcal{P}_o$ & The set of $p_o$ \\
\hline
$N_{o}$ & The maximum size of $\mathcal{P}_o$ \\
\hline
$u_p$ / $u_n$ & The parent / non-tree neighbor of $u$ in \cst \\
\hline
$t_{v}$ / $t_{n}$ & a visited / edge validation task \\
\hline
$\mathcal{T}_v$ / $\mathcal{T}_n$ & The set of $t_{v}$ / $t_{n}$ \\
\hline
\end{tabular}
\label{notations}
\end{center}
\controlspace
\controlspace
\controlspace
\end{table}

\subsection{Basic Pipeline of Subgraph Matching}
\label{hardware_implememtation_basic}

In the typical backtracking algorithms \cite{han2013turboiso,bi2016efficient,bhattarai2019ceci,han2019efficient}, one partial result is expanded at a time by matching the next vertex to a candidate vertex following the matching order. This sequential design cannot be pipelined because of data dependencies among iterations. To solve this, we decompose the matching process into three steps as follows: (1) \textit{Generator} expands partial results by matching the next vertex in the matching order; (2) \textit{Validator} verifies whether a new partial result is valid; (3) \textit{Synchronizer} collects results. Different from the typical algorithms, our method processes thousands of partial results at a time in these steps, so that each step can fully utilize the pipeline mechanism of FPGA. Our basic pipeline design is shown in Algorithm \ref{alg_fast_basic}, denoted as \textbf{FAST}.

\controlspace
\begin{algorithm}[htbp]  
\small
\caption{FAST($\cst, \mathcal{O}$)} \label{alg_fast_basic}
\KwIn{$\cst, \mathcal{O}$}
\KwOut{$\mathcal{M}$}
    $\mathcal{M}\leftarrow \emptyset$; $\mathcal{P}\leftarrow \emptyset$\; 
    \ForEach {candidate $v$ of root vertex \textbf{pipeline}}{\label{alg_fast_root_start}
        $\mathcal{P}.push(\{v\})$\;
    }\label{alg_fast_root_end}
    \BlankLine
    \While{$\mathcal{P} \neq \emptyset$}{\label{alg_fast_while}
        $\mathcal{P}_o,\mathcal{T}_{v},\mathcal{T}_{n}\leftarrow Generator(\mathcal{P}, \cst, \mathcal{O})$\;\label{alg_fast_generate}
        $\mathcal{B}_v\leftarrow VisitedValidator(\mathcal{T}_v$)\;\label{alg_fast_visit}
        $\mathcal{B}_n\leftarrow EdgeValidator(\cst,\mathcal{T}_n)$\;\label{alg_fast_edge}
        $Synchronizer(\mathcal{M},\mathcal{P},\mathcal{P}_o,\mathcal{B}_v,\mathcal{B}_n)$\;\label{alg_fast_syn}
    }
    \Return{$\mathcal{M}$}
\end{algorithm}
\controlspace
\controlspace

\begin{figure*}[htbp]
\centering
\controlspace
\subfigure[Basic pipeline of subgraph matching] {\includegraphics[width=0.55\columnwidth]{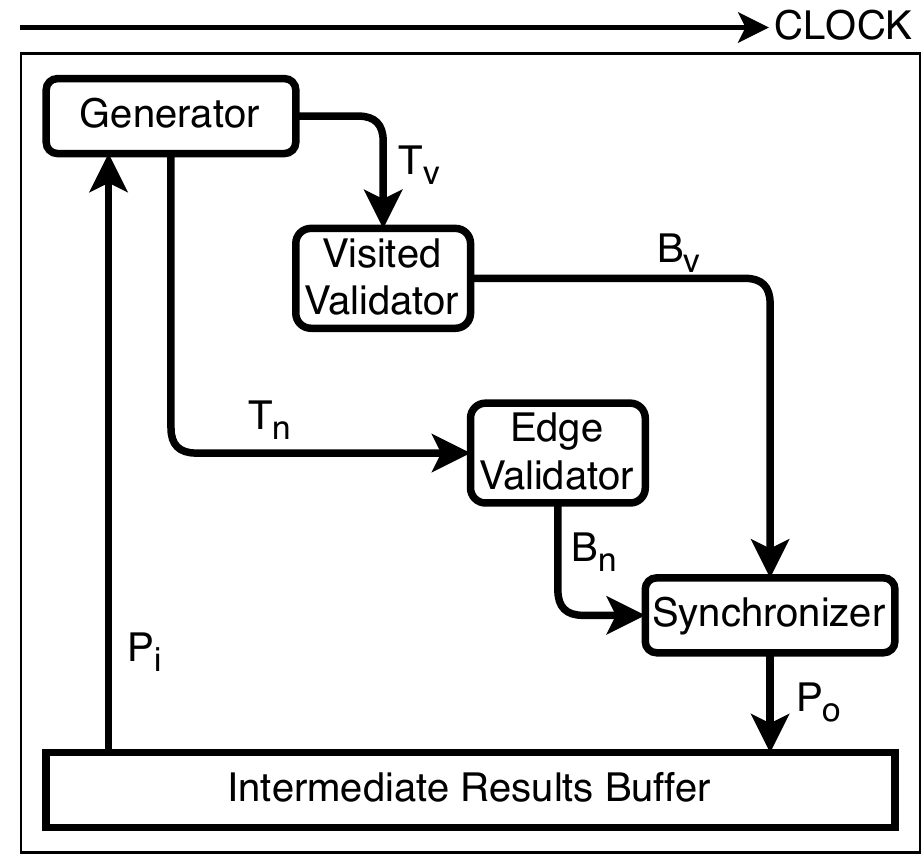}\label{fig_basic}}
\hspace{0.07\columnwidth}
\subfigure[Optimization with task parallelism]{
\includegraphics[width=0.55\columnwidth]{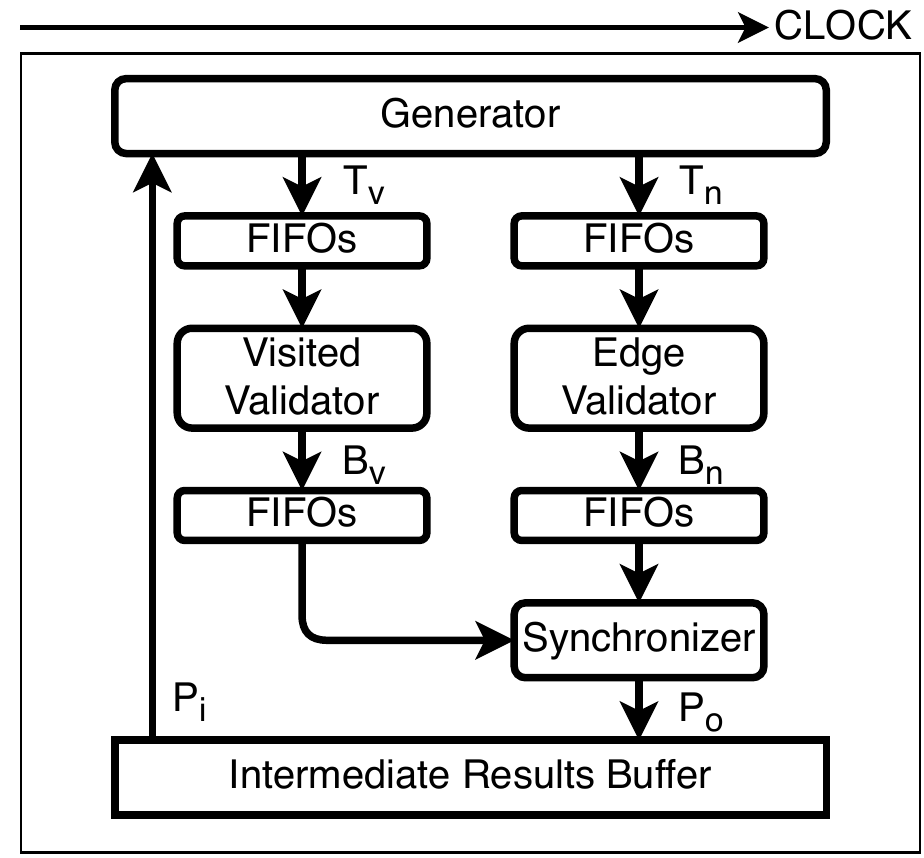}\label{fig_dataflow}}
\hspace{0.07\columnwidth}
\subfigure[Optimization for \emph{Generator}]{\includegraphics[width=0.55\columnwidth]{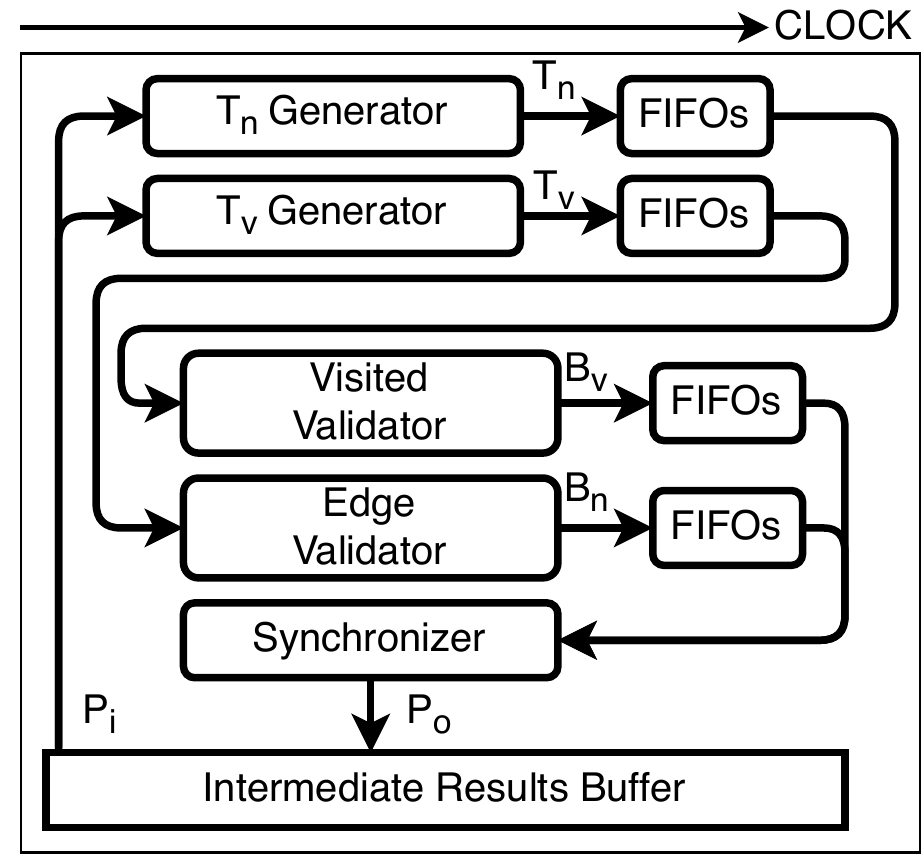}\label{fig_generator}}
\caption{The Hardware Implementation of \fast}
\label{fig_test_architecture}
\controlspace
\controlspace
\end{figure*}

Given \cst and matching order $\mathcal{O}$, we first match the root vertex to all its candidates to generate first batch of partial results (Line \ref{alg_fast_root_start}-\ref{alg_fast_root_end}). Then for each round, \emph{Generator} reads multiple partial results from $\mathcal{P}$ and expand them (Line \ref{alg_fast_generate}). A partial result is valid iff it passes the two validations: (1) \emph{visited validation}, i.e., the new mapped candidate $v$ is not visited before (Line \ref{alg_fast_visit}); (2) \emph{edge validation}, i.e., the new mapped candidate $v$ are adjacent to the mapping vertices of $u$'s non-tree neighbors (Line \ref{alg_fast_edge}). The new valid partial or complete results will be pushed into $\mathcal{P}$ or $\mathcal{M}$ by \emph{Synchronizer}, respectively (Line \ref{alg_fast_syn}). \textbf{FAST} terminates when $\mathcal{P}$ is empty (Line \ref{alg_fast_while}). As shown in Fig. \ref{fig_basic}, these steps are processed serially in our basic pipeline design. We discuss the details as follows.

\controlspace
\begin{algorithm}[htbp]  
\small
\caption{Generator($\mathcal{P},\cst,\mathcal{O}$)} \label{task_generator}
\KwIn{$\mathcal{P},\cst,\mathcal{O}$}
\KwOut{$\mathcal{P}_o,\mathcal{T}_v,\mathcal{T}_n$}
$\mathcal{P}_o\leftarrow \emptyset$; $\mathcal{T}_v\leftarrow \emptyset$; $\mathcal{T}_n\leftarrow \emptyset$\;
$u\leftarrow get\ next\ vertex\ to\ be\ mapped\ in\ \mathcal{O}$\;
\BlankLine
\tcc{Line \ref{alg_basic_loop1_start}-\ref{alg_basic_loop1_end}: Generate $\mathcal{P}_o$ and $\mathcal{T}_v$}
\While{$|\mathcal{P}_o|<{N}_o$}{\label{alg_basic_loop1_start}
    $p_i\leftarrow \mathcal{P}.pop()$\;
    $C(u)\leftarrow get\ u's\ candidates\ from\ \cst\ based\ on\ p_i$\;
    \lIf{$|\mathcal{P}_o|+ |C(u)|>N_o$}{\label{alg_max_out_start}break}
    \ForEach{$v\in C(u)$ \textbf{pipeline}}{
        $\mathcal{P}_o.push(p_i\times \{v\})$\;
        $\mathcal{T}_v.push((v,\ p_i))$\;\label{alg_generator_tv}
    }
    }\label{alg_basic_loop1_end}
\BlankLine
\tcc{Line \ref{alg_basic_loop3_outer_start}-\ref{alg_basic_loop3_outer_end}: Generate $\mathcal{T}_n$}
\ForEach{u's non-tree neighbor $u_n$}{\label{alg_basic_loop3_outer_start}
    \ForEach{$p_o\in \mathcal{P}_o$ \textbf{pipeline}}{ \label{alg_basic_loop3_inner_start}
    $\mathcal{T}_n.push(M_{p_o}(u),\ M_{p_o}(u_n),\ the\ index\ of\ p_o)$\;\label{alg_generator_tn}
    }\label{alg_basic_loop3_inner_end}
    }\label{alg_basic_loop3_outer_end}
\BlankLine
\Return{$\mathcal{P}_o,\mathcal{T}_v,\mathcal{T}_n$}
\end{algorithm}

\stitle{Generator.} \emph{Generator} is used to expand partial results and generate \emph{visited validation} tasks $\mathcal{T}_{v}$ and \emph{edge validation} tasks $\mathcal{T}_{n}$. Algorithm \ref{task_generator} shows the workflow of \emph{Generator}. At first, we expand partial results and generate visited validation tasks $\mathcal{T}_v$ (Line \ref{alg_basic_loop1_start}-\ref{alg_basic_loop1_end}). This procedure can be fully pipelined. Limited by on-chip resources, we control the maximum number of newly expanded partial results each round, denoted as $N_o$ (line \ref{alg_max_out_start}). We will discuss how to pick the value of $N_o$ in detail in Section \ref{cycle_analysis}. Then we generate edge validation tasks $\mathcal{T}_n$ (Line \ref{alg_basic_loop3_outer_start}-\ref{alg_basic_loop3_outer_end}). The inner loop of $t_n$ generation procedure is fully pipelined (Line \ref{alg_basic_loop3_inner_start}-\ref{alg_basic_loop3_inner_end}). We have specific one visited validation task $t_v$ for each new partial result $p_{o}$, while the number of edge validation tasks $t_{n}$ is determined by the query structure and matching order. One precondition to pipeline a loop is that the cycles of loop body are fixed. So we have to separate $\mathcal{T}_n$ generation procedure from other two steps. The outer loop of $\mathcal{T}_n$ generation procedure (Line \ref{alg_basic_loop3_outer_start}) cannot be pipelined for the same reason.

\controlspace
\begin{algorithm}[htb]  
\small
\caption{VisitedValidator($\mathcal{T}_v$)} 
\label{alg_visited_validator}
\KwIn{$\mathcal{T}_v$}
\KwOut{$\mathcal{B}_v$}
    $\mathcal{B}_v \leftarrow \emptyset$\;
    \ForEach{$(v, p_i)$ in $\mathcal{T}_{v}$\ \textbf{pipeline}}{\label{alg_fast_visit_for_start}
        $b\leftarrow 1$\;\label{alg_visited_fetch}
        \ForEach{$v'$ in $p_i$ \textbf{parallel}}{\label{alg_visited_start}
            \lIf{$v' == v$}{$b\leftarrow b\ \&\ 0$}
        }\label{alg_visited_end}
        $\mathcal{B}_{v}.push(b)$\;\label{alg_fast_visited_b}
    }\label{alg_fast_visit_for_end}
    \Return{$\mathcal{B}_v$}
\end{algorithm}
\controlspace
\controlspace

\stitle{Visited Validator.} As shown in Algorithm \ref{alg_visited_validator}, \emph{Visited Validator} is used to validate if the new mapped candidate $v$ is visited before by comparing $v$ with every vertex in $p_i$ (Line \ref{alg_visited_start}-\ref{alg_visited_end}). We use the array partition mechanism in FPGAs, i.e., partitioning an array into individual elements, to effectively increases the amount of read and write ports for the storage. The mechanism offers the possibility to compare $v$ with every element of $p_o$ in parallel. Each $p_o$ has two bits to reflect whether it passes visited and edge validation, respectively. If $v$ has been visited, the visited bit is set to zero (Line \ref{alg_fast_visited_b}). This module can be pipelined completely.

\controlspace
\begin{algorithm}[htbp]  
\small
\caption{EdgeValidator($\cst, \mathcal{T}_n$)} 
\label{alg_edge_validator}
\KwIn{\cst, $\mathcal{T}_n$}
\KwOut{$\mathcal{B}_n$}
    $\mathcal{B}_n \leftarrow \emptyset$\;
    \For{$(v, v_n, i)$ in $\mathcal{T}_{n}$\ \textbf{pipeline}}{\label{alg_fast_nt_for_start}
        \lIf{$(v, v_n)$ exists in \cst}{$b\leftarrow 1$}\label{alg_edge_existence}
        \lElse{ $b\leftarrow 0$ }\label{alg_edge_existence_failed}
        $\mathcal{B}_{n}.set(i,b)$\;
    }\label{alg_fast_nt_for_end}
    \Return{$\mathcal{B}_n$}
\end{algorithm}
\controlspace
\controlspace

\stitle{Edge Validator.} As shown in Algorithm \ref{alg_edge_validator}, \emph{Edge Validator} checks whether the new mapped candidate $v$ is adjacent to all $v_n$, the mappings of $u$'s non-tree neighbors. It checks edge existence in \cst (Line \ref{alg_edge_existence}) by comparing $v_n$ with all non-tree candidate neighbors of $v$. Here we also adopts the array partition mechanism so that edge existence check can be completed in $O(1)$. However, this mechanism costs much more on-chip resources, which limits the maximum number of access ports of an array, denoted as $Port_{max}$. Thus we partition \cst if $D_{\cst}$ exceeds $Port_{max}$. If there is no edge between $v$ and $v_n$, the edge bit is set to zero (Line \ref{alg_edge_existence_failed}). It should be noted that each $p_o$ may have more than one $t_n$, any of them failed will lead to an invalid $p_o$. The \emph{Edge Validator} module can also been pipelined completely.

\controlspace
\begin{algorithm}[htb]  
\small
\caption{Synchronizer($\mathcal{M},\mathcal{P},\mathcal{P}_o,\mathcal{B}_v,\mathcal{B}_n$)} 
\label{alg_synchronizer}
\KwIn{$\mathcal{M},\mathcal{P},\mathcal{P}_o,\mathcal{B}_v,\mathcal{B}_n$}
    \For{$p_o$ in $\mathcal{P}_o$\ \textbf{pipeline}}{\label{alg_fast_sy_start}
        $b_v\leftarrow \mathcal{B}_v.pop(), b_n\leftarrow \mathcal{B}_n.pop()$\;\label{alg_syn_fetch}
        \If{$b_v=1$ and $b_n=1$}{\label{alg_syn_check_bit}
            \lIf{$|p_o| == |O|$}{\label{alg_syn_complete}
                $\mathcal{M}.push(p_o)$
            }\lElse{
                $\mathcal{P}.push(p_o)$
            }
        }
    }
\end{algorithm}
\controlspace
\controlspace

\stitle{Synchronizer.} As shown in Algorithm \ref{alg_synchronizer}, \emph{Synchronizer} is designed to collect partial results. For each $p_o$, it first fetches its two validation bits from $\mathcal{B}_v$ and $\mathcal{B}_n$ (Line \ref{alg_syn_fetch}). If any bit is zero, this $p_o$ will be discarded (Line \ref{alg_syn_check_bit}). Then it compares $|p_o|$ and $|\mathcal{O}|$ to check whether it is a complete result (Line \ref{alg_syn_complete}). The complete result is reported and stored into $\mathcal{M}$ while the partial result is stored back into $\mathcal{P}$. 

\begin{example}
Suppose that we have \cst in Fig. \ref{fig_partition_a},  $\mathcal{O}=(u_0,u_1,u_2,u_3)$ and $\mathcal{P}=\{\{v_1,v_3\},\{v_1,v_5\}\}$. The \emph{Generator} first expand partial results in $\mathcal{P}$ to get $\mathcal{P}_o=\{\{v_1,v_3,v_6\},\{v_1,v_3,v_8\},\{v_1,v_5,v_6\},\{v_1,v_5,v_8\}\}$ and generate $\mathcal{T}_v=\{(v_6,0),(v_8,0),(v_6,1),(v_8,1)\}$. After that, $\mathcal{T}_n=\{(v_3,v_6,0),(v_3,v_8,1),(v_5,v_6,2),(v_5,v_8,3)\}$ are generated. Then \emph{Visited Validator} and \emph{Edge Validator} processes $\mathcal{T}_v$ and  $\mathcal{T}_n$, respectively. We get $\mathcal{B}_v=\{1,1,1,1\}$ and $\mathcal{B}_n=\{1,0,0,1\}$. Finally, \emph{Synchronizer} pushes valid partial results $\{\{v_1,v_3,v_6\},\{v_1,v_5,v_8\}\}$ into $\mathcal{P}$.
\end{example}

\subsection{Cycle Analysis and Buffer Design}
\label{cycle_analysis}
In this subsection, we first discuss how to pick the value of the maximum number of newly expanded partial results each round, denoted as $N_o$. Then our BRAM-only intermediate results buffer is introduced, which completely avoids the intermediate data transfer between BRAM and DRAM.


Based on Algorithm \ref{task_generator}-\ref{alg_synchronizer}, we use $L_1-L_6$ to denote the average cycles for the following six procedures: (1) read from intermediate results buffer $\mathcal{P}$; (2) generate a new partial result $p_o$ and its visited validation task $t_v$; (3) process $t_v$; (4) collect $p_o$; (5) generate an edge validation task $t_n$; (6) process $t_n$. We use $m$ to denote the number of $t_n$ for $p_o$. So the total cycles of a partial result from being expanded to finally collected are $(L_1+L_2+L_3+L_4+n\times (L_5+L_6))$.

Suppose in the whole search space, the total number of $p_o$ and $t_n$ is $N$ and $M$, respectively. To simplify the equations, we denote $\sum_{j=1}^{4}L_i$ as $L_f$ and $\sum_{j=5}^{6}L_i$ as $L_t$. So without any pipelining optimization, the total cycles $L_{serial}$ to process the whole search space is:
\begin{equation}
\small
L_{serial}=N\times L_f+M\times L_t
\end{equation}

In \fast, the six procedures can be pipelined completely and we process $N_o$ partial results each round. It means each round the serial algorithm needs $L_2\times N_o$ cycles to process the second procedure while \fast needs $(L_2+N_o+1)$ cycles. So the total cycles $L_{basic}$ is:
\begin{equation}
\label{equ_basic}
\small
L_{basic}\approx\frac{N\times L_f+M\times L_t}{N_o}+4N+2M
\end{equation}

As shown in Equation \ref{equ_basic}, a small $N_o$ decreases the performance. However, it leads to over-consumption of on-chip resources when $N_o$ is too large. Thus we ensure $N_o>>\frac{N\times L_f+M\times L_t}{4N+2M}$ and the specific value of $N_o$ should be carefully chosen based on different FPGAs (our configuration is given in Section \ref{experiments}). It should be noted that for a partial result, if its candidates are too many, i.e., $|C(u)|>N_o$, we will generate $N_o$ partial results by mapping $N_o$ candidates in $C(u)$. The rest candidates will be mapped later.

It is expensive to transfer partial results between BRAM and DRAM. So we develop a strategy to avoid the overflow of the intermediate results buffer $\mathcal{P}$. We use $p^n$ to denote a partial result that maps $n$ query vertices. We observe that a $p^{|V(q)|}$ is a complete result and will not be pushed back into $\mathcal{P}$. Therefore, each round we expand $p^n$ with the maximum $n$ in $\mathcal{P} $so that these partial results can be expanded to complete ones as soon as possible. As a result, for any $n\in[1,|V(q)|-1]$, our strategy guarantees the number of $p^n$ does not exceed $N_o$. Finally, we allocate $(|V(q)|-1)\times N_o$ space for $\mathcal{P}$ on BRAM, which prevents the overflow of $\mathcal{P}$. 
\subsection{Optimization with Task Parallelism}
\label{dataflow}
As shown in Fig. \ref{fig_basic}, in our basic pipeline design, modules are executed in serial. The number of access ports to ordinary memory area on BRAM is limited, so two modules can not access the same memory simultaneously. As a result, \emph{Visited Validator} and \emph{Edge Validator} cannot start until all $t_v$ and $t_n$ are generated. \emph{Synchronizer} will be idle before all validation tasks are finished. Therefore, as illustrated in Fig. \ref{fig_dataflow}, we utilize \emph{task parallelism} mechanism on FPGA to allow modules being executed in parallel.

In contrast to loop parallelism, when task parallelism is deployed, different execution modules are allowed to operate simultaneously. The task parallelism is achieved by taking advantage of extra buffering introduced between the modules. The buffer is implemented by FIFOs (First in, First out) on FPGA. The output of each module will be streamed into the buffer, and the next module processes the data as long as the buffer is not empty. 

As shown in Algorithm \ref{task_generator}, once $t_v$ is generated (Line \ref{alg_generator_tv}), it is streamed to the FIFOs, and \emph{Visited Validator} starts to work. Similarly, once $t_n$ is generated (Line \ref{alg_generator_tn}), it is streamed to the FIFOs, and \emph{Edge Validator} starts its process. The $p_o$ will be collected by \emph{Synchronizer} as soon as its two validation bits are ready. Compared with the basic version, more than one module can work simultaneously.

Consider the total cycles of this task parallelism version, denoted as $L_{task}$. In this optimized design, the first loop of \emph{Generator} (Line \ref{alg_basic_loop1_start}-\ref{alg_basic_loop1_end} in Algorithm \ref{task_generator}) and \emph{Visited Validator} (Algorithm \ref{alg_visited_validator}) execute in parallel. And the second loop of \emph{Generator} (Line \ref{alg_basic_loop3_outer_start}-\ref{alg_basic_loop3_outer_end} in Algorithm \ref{task_generator}), \emph{Edge Validator} (Algorithm \ref{alg_edge_validator}) and \emph{Synchronizer} (Algorithm \ref{alg_synchronizer}) execute concurrently. To simplify the equation, suppose we have pick an appropriate $N_o$. Then we have:
\begin{equation}
\label{equ_stream}
    \small
    L_{task}\approx2N+\max(N,M)
\end{equation}

Compared with Equation \ref{equ_basic}, this optimization can achieve up to $50\%$ performance improvement in theory.
\subsection{Optimization for Generator}
\label{hardware_generator}

As shown in Fig. \ref{fig_dataflow}, in our task parallelism version, $t_n$ generation procedure has to wait until $t_v$ generation procedure finishes, which decreases the overall throughput. \emph{Synchronizer} also waits for the output of \emph{Edge Validator}, although all visited bits of $p_o$ are ready. Therefor, we carry out optimizations on \emph{Generator} module.

\emph{Generator} module is split into \emph{$t_v$ Generator} and \emph{$t_n$ Generator}. Once a new $p_o$ is generated, it will be copied so that the source and the copy of $p_o$ can be streamed into different FIFOs of two generators separately. Both $t_v$ Generator and $t_n$ Generator can start to work while \emph{Synchronizer} starts to collect partial results at the same time. This optimization is achieved by copying data and using more on-chip resources (e.g., FIFOs). Thanks to the loop parallelism characteristic of FPGA, the cost of copy of $p_o$ does not decrease the performance. And we analyze the total cycles of this optimized version $L_{sep}$. All modules execute concurrently. As a result, the minimum cycles we can achieve is as follows:
\begin{equation}
\label{equ_sep}
    \small
    L_{sep}\approx N+\max(N,M)
\end{equation}

Compared with equation \ref{equ_stream}, this optimization can achieve at most $33\%$ performance improvement theoretically.
\section{Experiments}
\label{experiments}
We present the results of our performance studies in this section. We first introduce the experimental setup of the experiments. Then, we investigate the necessity of \cst partition 
and evaluate the effectiveness of our software and hardware optimizations, followed by the comparison with the state-of-the-art algorithms. We also evaluate our algorithm on a billion-scale graph to test the scalability.

\stitle{Algorithms.} We compare two state-of-the-art GPU-based solutions: \gsi\cite{zeng2020gsi} and \gpsm\cite{tran2015fast}. According to the latest survey \cite{sun2020memory}, we compare other three state-of-the-art CPU-based algorithms: \cfl \cite{bi2016efficient}, \daf \cite{han2019efficient}, \ceci \cite{bhattarai2019ceci} and five versions of our algorithm:
\begin{itemize}[leftmargin=*]
    \item \emph{FAST-DRAM}: the algorithm fetches data from DRAM without any other optimizations.
    \item \emph{FAST-BASIC}: the algorithm fetches data from on-chip memory without any other optimizations (Section \ref{hardware_implememtation_basic}).
    \item \emph{FAST-TASK}: \emph{FAST-BASIC} algorithm boosted by the task parallelism optimization (Section \ref{dataflow}).
    \item \emph{FAST-SEP}: \emph{FAST-BASIC} algorithm boosted by the both task parallelism and task generator separation optimization (Section \ref{hardware_generator}).
    \item \emph{FAST-SHARE}: \emph{FAST-SEP} algorithm where the host side, i.e. CPU, shares some matching tasks (Section \ref{software_share}).
\end{itemize}

Among the five versions, we choose \emph{FAST-SHARE} as the final version of our algorithm, denoted as \fast.
The parallel version of \daf and \ceci are also evaluated, denoted as \daf-8 and \ceci-8 respectively, which run on 8 CPU threads. For all other algorithms, we use only one CPU thread.

\stitle{Setup.} We implement \fast in C++ on an Alveo U200 Data Center Accelerator Card, equipped with 64GB off-chip DRAM, 35MB on-chip BRAM, and communicates with the host through PCIe gen3 $\times$ 16. It runs at 300 MHz on the FPGA card.
All experiments are conducted on a machine equipped with an 8-core Intel Xeon E5-2620 v4 CPU (2.1GHz), 250G host memory, NVIDIA Tesla V100 (5120 streaming processors, 16GB global memory), running Ubuntu 16.04. 

\begin{table}[htbp]
\controlspace
\controlspace
\caption{Characteristics of datasets.}
\begin{center}
\begin{tabular}{|c|c|c|c|c|c|}
\hline
Name & $|V_G|$ & $|E_G|$ & $\Bar{d_G}$ & $D_G$ & $\#$ Labels\\
\hline
DG01 & 3.18M & 17.24M & 10.84 & 464,368 & 11\\
\hline
DG03 & 9.28M & 52.65M & 11.34 & 1,346,287 & 11\\
\hline
DG10 & 29.99M & 176.48M & 11.77 & 4,282,812 & 11\\
\hline
DG60 & 187.11M & 1.25B & 13.33 & 26,639,563 & 11\\
\hline
\end{tabular}
\label{tab_datasets}
\end{center}
\controlspace
\controlspace
\controlspace
\end{table}


\stitle{Datasets.} The datasets commonly used in previous works \cite{ren2015exploiting,bi2016efficient,han2019efficient,han2013turboiso} are composed of small-scale data graphs (e.g., Yeast with 3.11K vertices and 12.51K edges) and large queries (e.g., 200 vertices), whereas the data graphs are usually very large and the queries are relatively small in real-world workloads nowadays.
Therefore, we adopt the LDBC social network benchmarking (\emph{LDBC-SNB}) \cite{ldbcbenchmark} in our experiment to simulate real-world workloads.
The LDBC-SNB benchmark serves as an industry-standard benchmarking and provides a data generator that generates a synthetic social network together with a set of benchmarking tasks, in which many tasks are subgraph matchings.

We list the datasets and their statistics in Table \ref{tab_datasets}. These datasets are generated simulating a real social network akin to Facebook with a duration of 3 years. The dataset’s name, denoted as DG$x$, represents a scale factor of $x$.

\begin{figure}[htbp]
\controlspace
\centerline{\includegraphics[width=0.9\columnwidth]{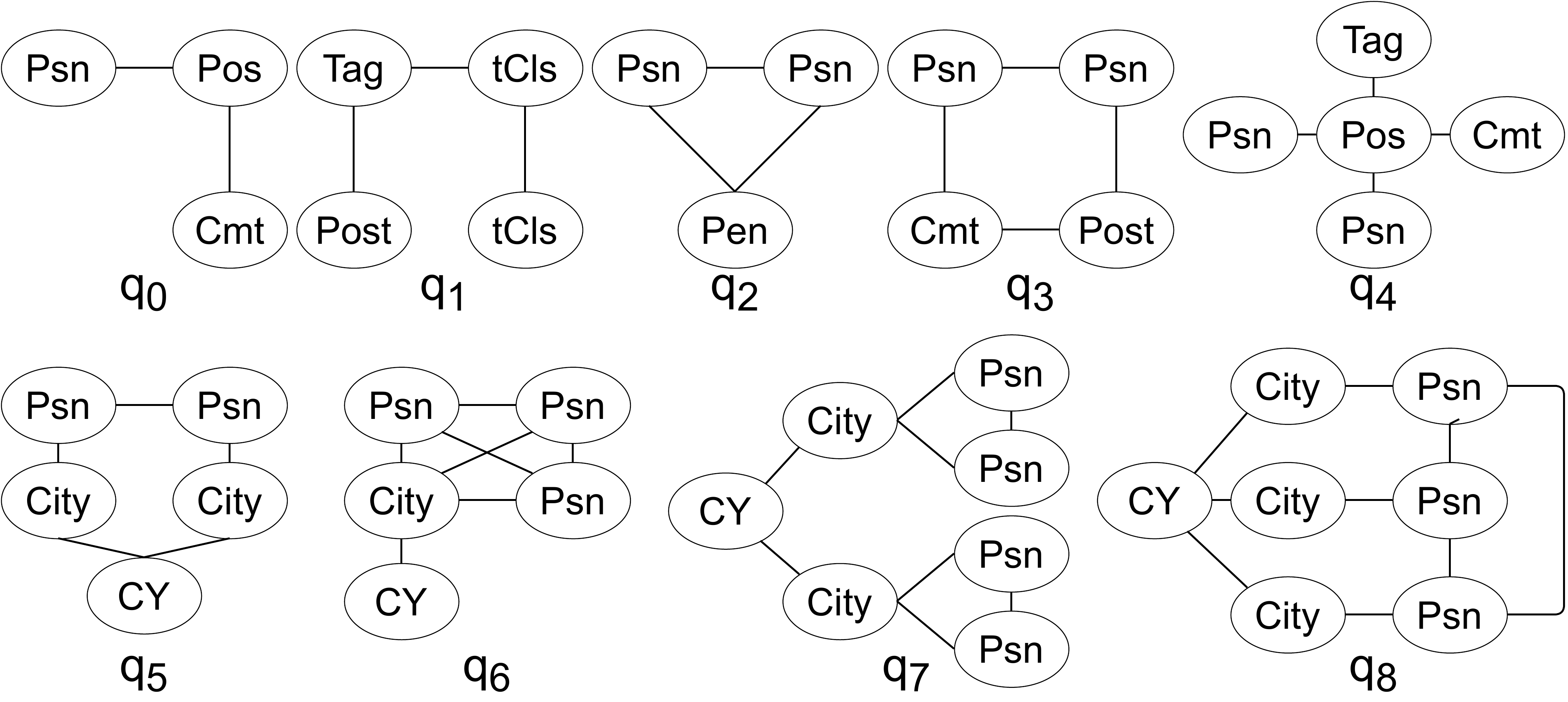}}
\controlspace
\caption{The Queries}
\label{fig_queries}
\controlspace
\end{figure}

\stitle{Queries.} We use the queries in \cite{lai2019distributed}, as shown in Fig. \ref{fig_queries}. The queries are selected from the LDBC-SNB's complex tasks with some adaptions, including only keeping the node types as labels and removing multi-hop edges in order to conform with the subgraph matching problem studied in this paper.

\stitle{Metrics.}
To evaluate an algorithm, we measure the execution time in milliseconds. We set a time limit of 3 hours for each query. Each query is run three times and the \emph{average} time is reported. We denote the execution time of queries with timeout as `INF' and queries running out of memory as `OOM'.

\subsection{The Necessity of \cst Partition}
\label{experiment_dram}
We partition \cst in order to store it in BRAM instead of DRAM on FPGA because of the much higher read latency of DRAM. On the other hand, the random read of \cst leads to the impossibility to prefetch the data from DRAM into BRAM. We compare the elapsed time of \emph{FAST-DRAM} and \emph{FAST-BASIC} to verify the necessity of \cst partition.

\begin{figure}[htbp]
\centerline{\includegraphics[width=0.95\columnwidth]{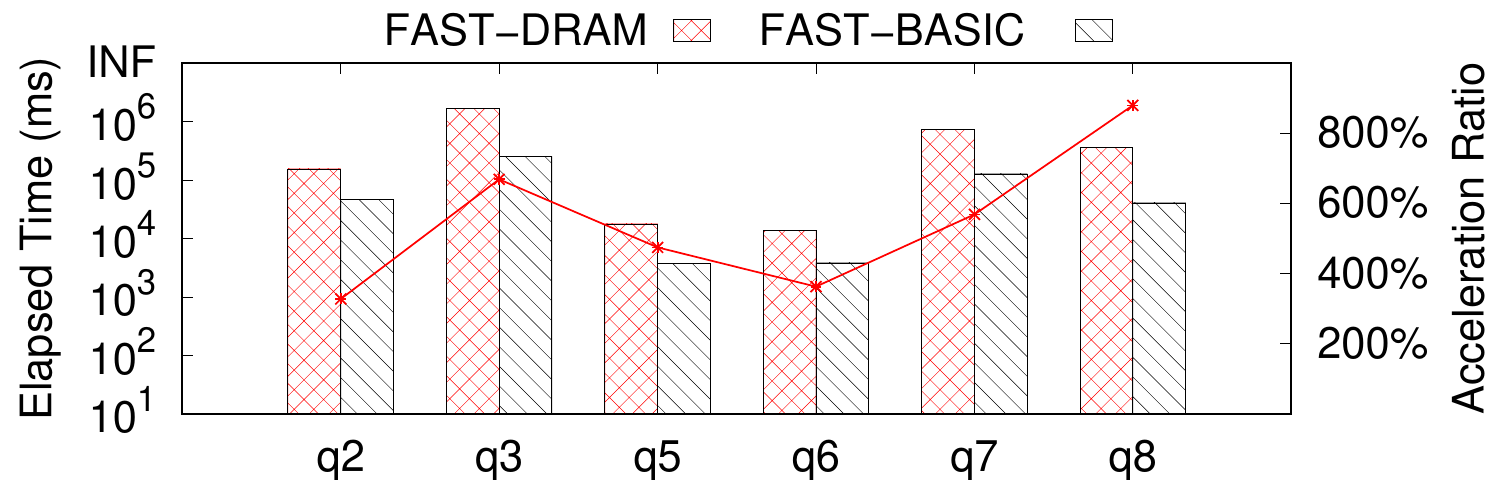}}
\caption{Elapsed time of \emph{FAST-DRAM} and \emph{FAST-BASIC} for DG10}
\label{fig_dram}
\end{figure}

As shown in Fig. \ref{fig_dram}, the results indicate that \emph{FAST-BASIC} outperforms \emph{FAST-DRAM} for all the queries in both DG03 and DG10. Despite the initial overhead to fetch data from DRAM to BRAM, \emph{FAST-BASIC} achieves about 5.0x speedup compared with \emph{FAST-DRAM} on average. The speedup is close to the ratio of the read latency. Moreover, it is confirmed by the growing speedup (4.50x for DG01, 5.18x for DG03, and 5.93x for DG10) that the initial transmission overhead has a decreasing impact on the overall performance for a larger graph. These results show the necessity to partition \cst structure to avoid direct data access to DRAM.

\begin{figure}[htbp]
\centerline{\includegraphics[width=1\columnwidth]{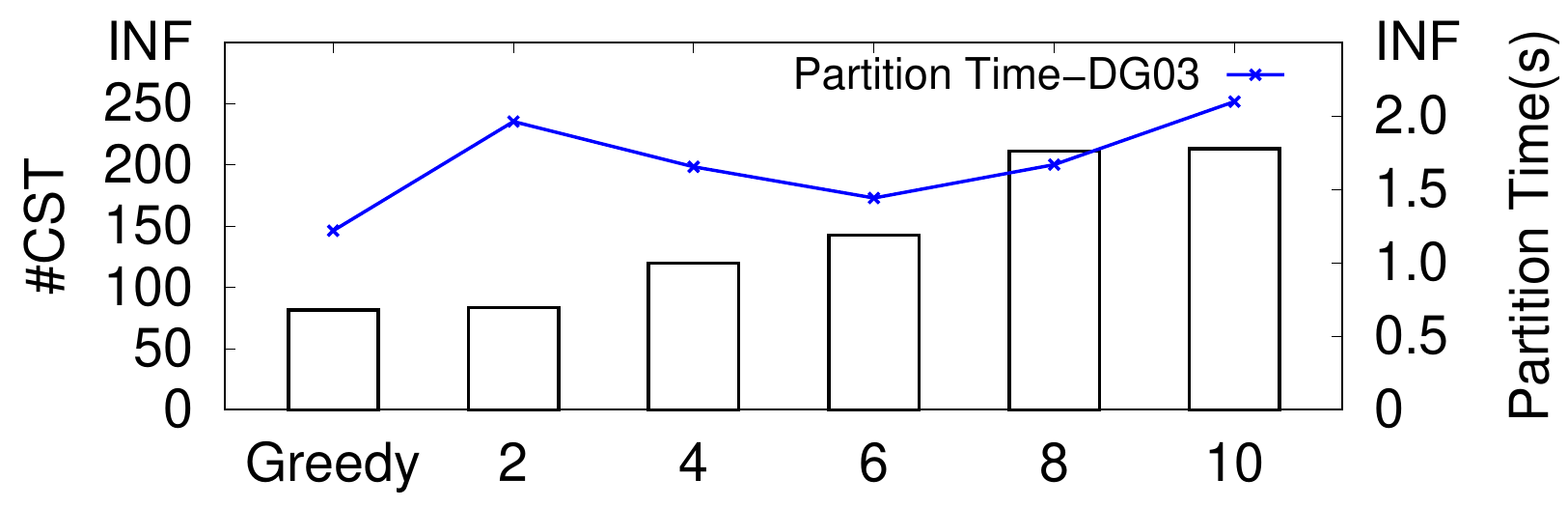}}
\caption{The average number of CST and average partition time varying $k$}
\label{fig_k}
\end{figure}

We have added the $k$-Determination experiment as illustrated in Fig. \ref{fig_k} to evaluate the impact of partition factor $k$. Besides our greedy strategy, we test \fast with fixed $k\in\{2,4,6,8,10\}$. The average number of CST and the average partition time are reported. It can be seen that our greedy approach does achieve the least number of CST and least time cost to partition CST. The acceleration is not particularly sensitive to $k$ when $k$ is small (e.g. $k \leq 10$). The choice of $k$ do make impact on the partition time, but when $k$ is small, the partition time of CPU is overlapped well with the time of computing matchings on FPGA (the more time-consuming process).
When $k$ is large, the partition time can potentially increase rapidly and hence harm the acceleration. However, our greedy strategy can select a good $k$ to reduce the time for partitioning and the final number of CST partitions, so it can make less impact on the whole subgraph matching process.

\begin{figure}[htbp]
\centerline{\includegraphics[width=0.95\columnwidth]{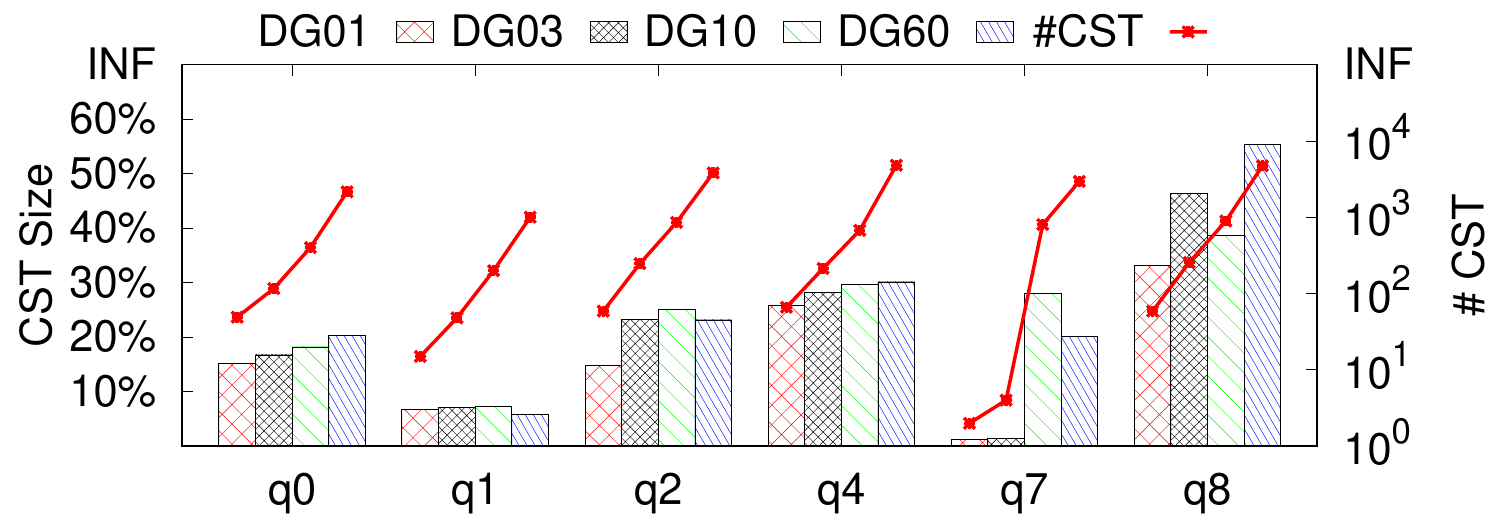}}
\caption{The number and total size of partitioned \cst}
\label{fig_cst_num}
\end{figure}

 We use $S_{\cst}$ and $S_{G}$ to denote the size of all \cst partitions and data graph, respectively. Fig. \ref{fig_cst_num} illustrates the number of \cst partitions and $\frac{S_{\cst}}{S_{G}}$, 
As expected, the number of \cst partitions increases for larger data graphs. And $\frac{S_{\cst}}{S_{G}}$ keeps stable for most queries while the data graph grows ($\frac{S_{\cst}}{S_{G}} < 60\%$ for all queries). The rapid growth of $\frac{S_{\cst}}{S_{G}}$ in $q_7$ from DG03 to DG10 is due to the rapid increase in the number of embeddings. 

\begin{figure}[htbp]
\centerline{\includegraphics[width=1\columnwidth]{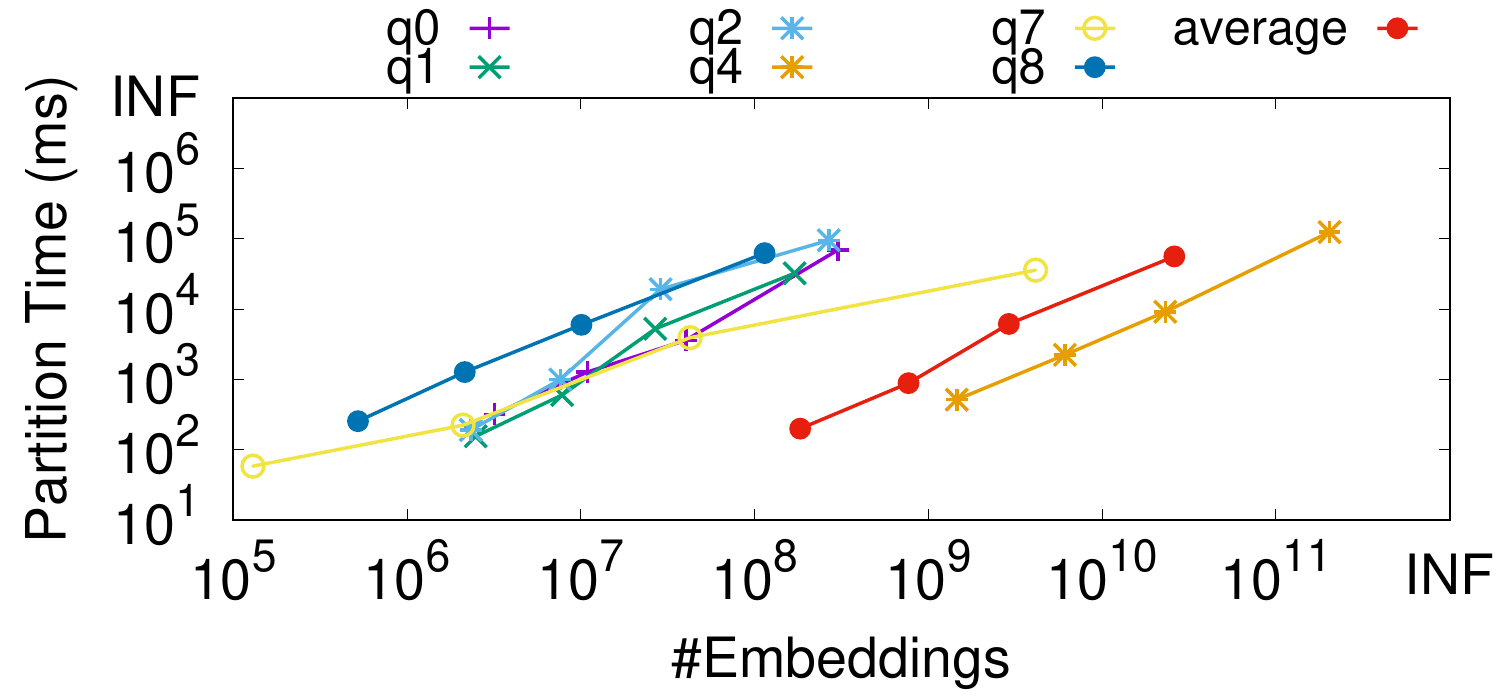}}
\caption{The partition time per embedding}
\label{fig_cst}
\end{figure}

Moreover, we evaluate the partition time with respect to the embeddings as the data graph grows. The results are shown in Fig. \ref{fig_cst}. The average partition time increases only slightly as data graph grows ($1.09\times10^{-9}$, $1.15\times10^{-9}$, $2.11\times10^{-9}$, and $2.15\times10^{-9}$ seconds per embedding for DG01, DG03, DG10 and DG60, respectively), while the sizes of data graph increase a lot (the numbers of edges are 17.24M, 52.65M, 176.48M and 1.25B for DG01, DG03, DG10 and DG60, respectively).
The results of memory cost and time cost proves the scalability of our partition mechanism when the data graph grows.

\subsection{Evaluating Optimization Techniques}
In this section, we test the four versions of our algorithm: \emph{FAST-BASIC}, \emph{FAST-TASK}, \emph{FAST-SEP} and \emph{FAST-SHARE} to evaluate the effectiveness of our software and hardware optimizations.

\begin{figure}[htbp]
\controlspace
\centerline{\includegraphics[width=0.95\columnwidth]{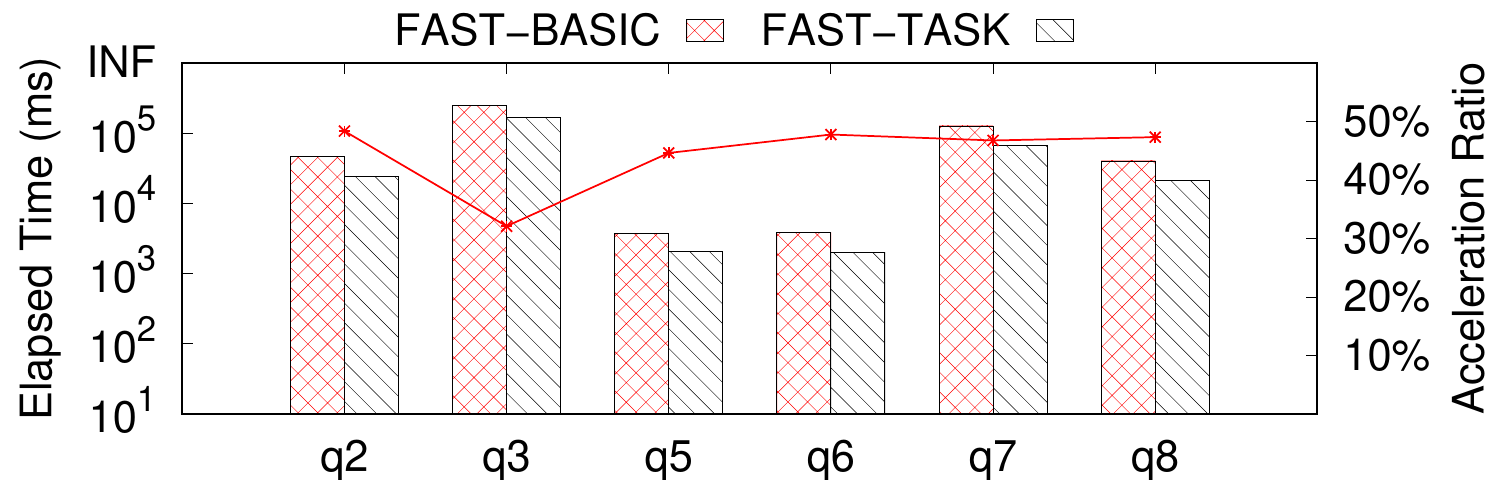}}
\controlspace
\caption{Elapsed time of \emph{FAST-BASIC} and \emph{FAST-TASK} for DG10}
\label{fig_opt_stream}
\controlspace
\end{figure}

\stitle{Effectiveness of \emph{Task Parallelism}}.
\emph{FAST-BASIC} only adopts the loop pipeline mechanism of FPGA, while \emph{FAST-TASK} introduces task parallelism so that full execution modules are allowed to operate in parallel. 
From the acceleration ratio in Fig. \ref{fig_opt_stream}, we can see that the task parallel optimization achieves up to $50\%$ improvement (e.g. $q_8$). The theoretical improvement of task parallelism is discussed in Section \ref{dataflow}. From Equation (\ref{equ_basic}) and Equation (\ref{equ_stream}), we can see that the task parallelism optimization achieves better performance for dense queries whose $M$ is larger than $N$. The acceleration ratio of $q_3$ is much lower than other queries because of its much higher $\frac{N}{M}$ (about $2$ for $q3$ and close to or lower than $1$ for other queries).

\begin{figure}[htbp]
\controlspace
\centerline{\includegraphics[width=0.95\columnwidth]{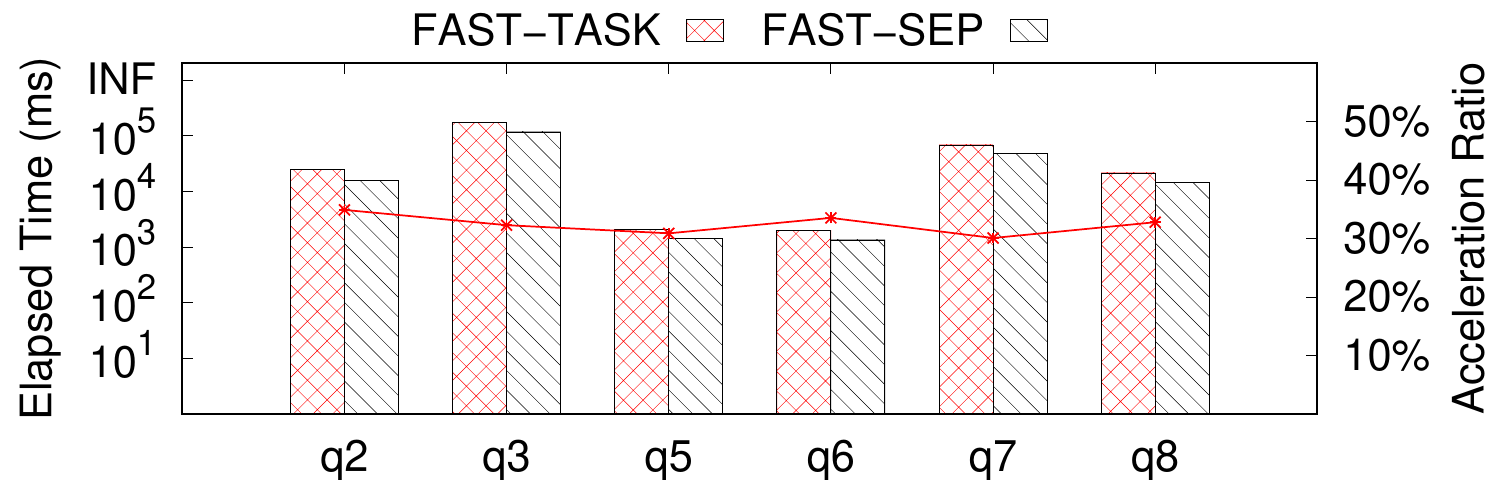}}
\controlspace
\caption{Elapsed time of \emph{FAST-TASK} and \emph{FAST-SEP} for DG10}
\label{fig_opt_separation}
\controlspace
\end{figure}

\stitle{Effectiveness of \emph{Task Generator Separation}}. The task parallelism allows all modules to execute concurrently. However, it is limited by the first module \emph{Generator} to generate two kinds of tasks in parallel so that the following modules can start to work at the same time. \emph{FAST-SEP} solves this problem by using more on-chip resources and duplicating data. Compared with the average elapsed time of \emph{FAST-TASK}, \emph{FAST-SEP} achieves about $30\%-40\%$ improvements (e.g. $q8$). The effectiveness of \emph{Task Generator Separation} is consistent with our cycle analysis in Equation \ref{equ_stream} and Equation \ref{equ_sep}. 
Moreover, when $\frac{N}{M}>1$, \emph{Task Generator Separation} achieves the best improvements.

\begin{figure}[htbp]
\controlspace
\centerline{\includegraphics[width=1\columnwidth]{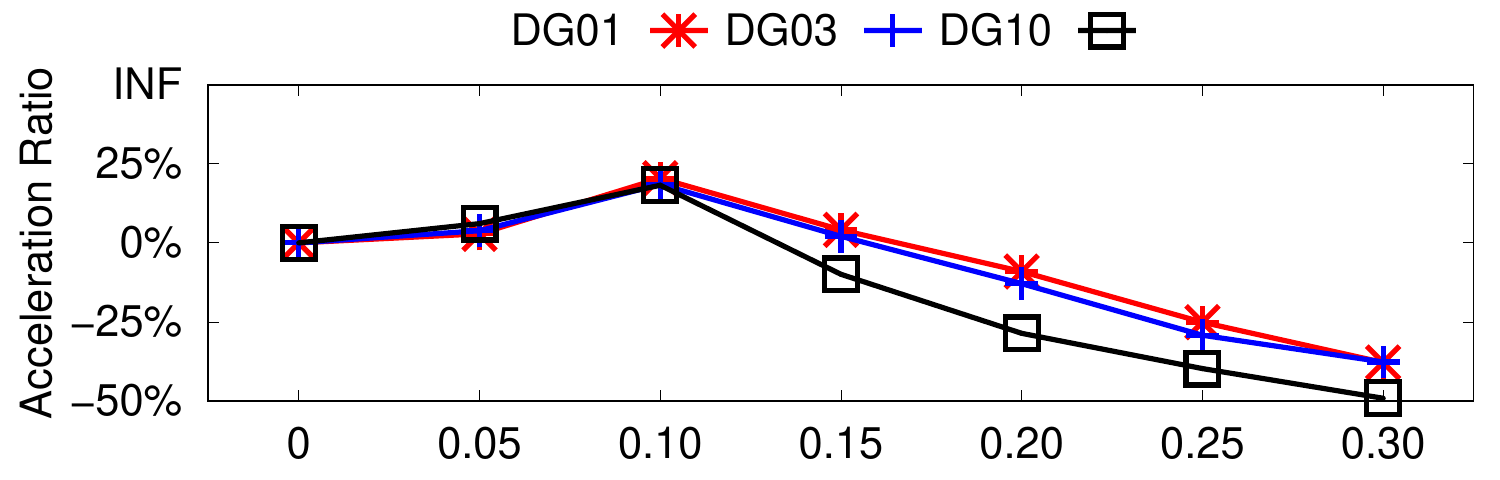}}
\controlspace
\caption{Average acceleration ratio varying $\delta$}
\label{fig_opt_cpu}
\controlspace
\end{figure}

\stitle{Effectiveness of \emph{Software Scheduler}}. After partitioning \cst, CPU becomes idle, which can be utilized to share some matching tasks. We propose a workload estimation method of \cst and restrict the proportion of the total workload of matching tasks assigned to CPU from exceeding a threshold $\delta$. We evaluate the effectiveness of \emph{software scheduler} by varying $\delta$. The results in Fig. \ref{fig_opt_cpu} indicate that this optimization achieves biggest improvements when $delta=0.1$ (e.g. $20\%$ for DG01). The reason that this optimization achieves more than $10\%$ improvements  ($\delta = 0.1$ which means the host side shares about $10\%$ matching tasks) is as follows: Considering \cst that can not be fully offloaded into BRAM, in \emph{FAST-SEP}, we have to partition it until it meets the size constraints; in \emph{FAST-SHARE}, we may directly assign it to CPU, reducing the cost of partitioning. Moreover, it can be seen from the figure that the CPU becomes the bottleneck when $delta > 0.15$.

\subsection{Comparing with Existing Algorithms}

\begin{figure*}[htbp]
\centering
\subfigure[DG01]{\label{fig_algorithm_1}
\includegraphics[width=0.66\columnwidth]{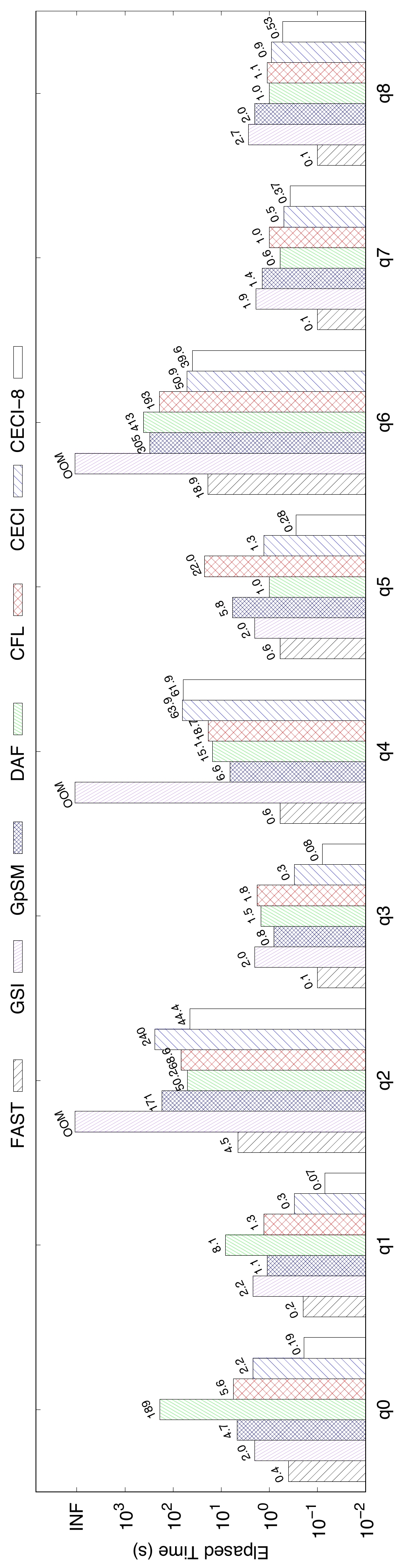}}
\subfigure[DG03]{\label{fig_algorithm_3}
\includegraphics[width=0.66\columnwidth]{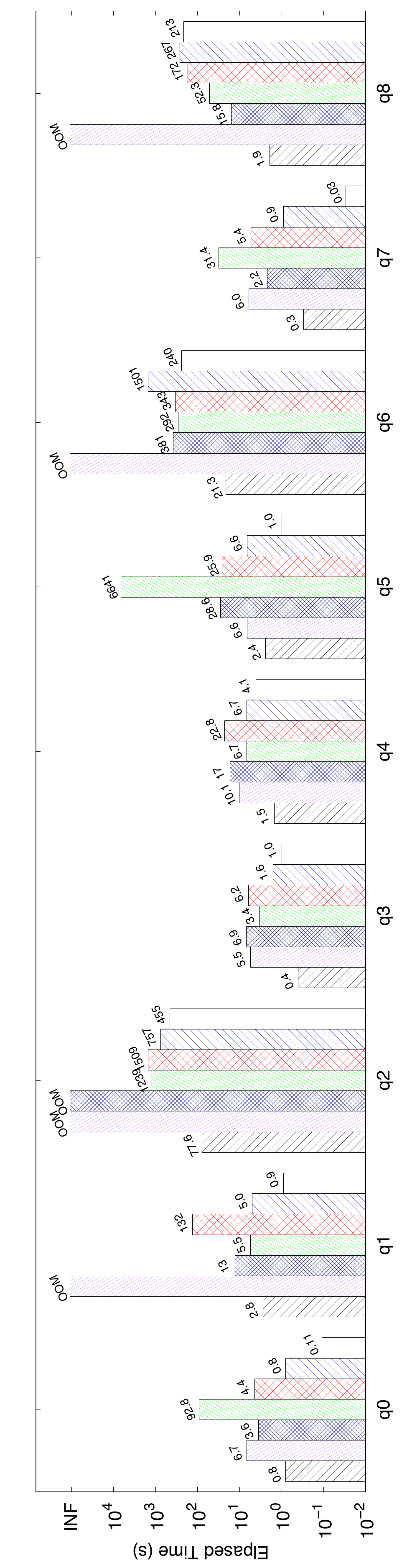}}
\subfigure[DG10]{\label{fig_algorithm_10}
\includegraphics[width=0.66\columnwidth]{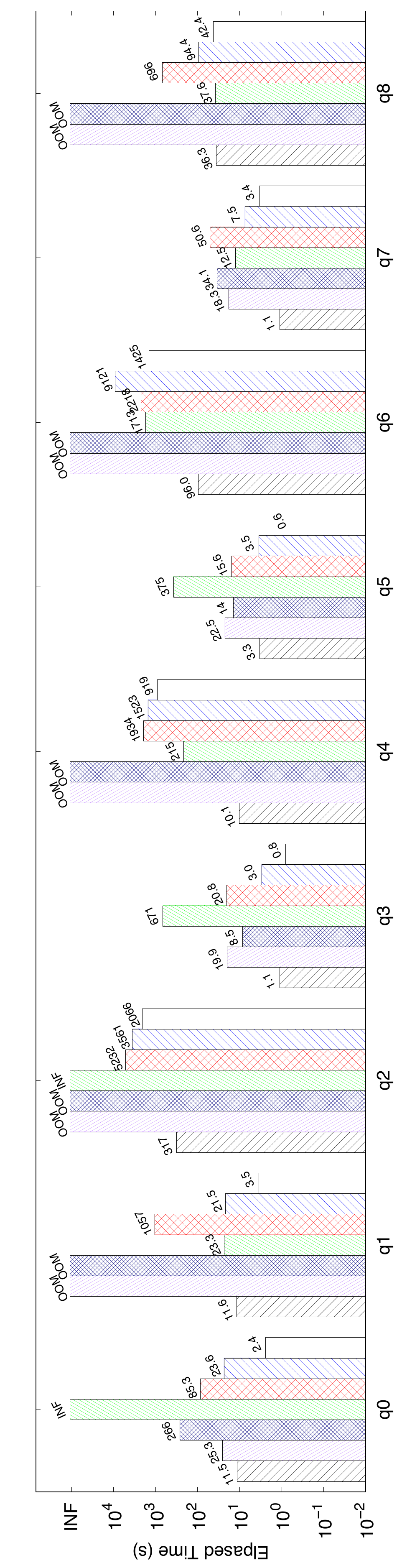}}
\caption{Elapsed time of CFL-Match, CECI, DAF and FAST}
\label{fig_algorithm}
\vspace{-1.2\baselineskip}
\end{figure*}

We then evaluate \fast against the existing algorithms, \gsi \cite{zeng2020gsi}, \gpsm \cite{tran2015fast}, \cfl \cite{bi2016efficient}, \daf \cite{han2019efficient} and \ceci \cite{bhattarai2019ceci}. 
All the source code comes from the original authors and is also implemented in C++. 
Fig \ref{fig_algorithm} shows the experimental results. Each query show similar trend in different data graphs, so we only demonstrate the results of five queries for each data graph due to the space limit.
\fast outperforms all other algorithms for all the queries and achieves 24.6x average speedup. Specifically, \fast outperforms \gsi by up to 36.6x ($q_6$ in DG01), outperforms \gpsm by up to 38.0x ($q_3$ in DG01), outperforms \cfl by up to 191.0x ($q_8$ in DG10), outperforms \daf by up to 462.0x ($q_0$ in DG01) and outperforms \ceci by up to 150.0x ($q_8$ in DG10). 

We noticed that the GPU-based solutions do not show better performance over CPU-based algorithms for some queries. More critically, both \gsi and \gpsm are only able to handle the graphs that can be fit into the GPU memory. So they both fail to solve all the queries. The reason why \gsi has a higher memory cost is that \gsi pre-allocates enough memory space instead of joining twice like \gpsm to avoid the conflicts when each processor writes results to memory in parallel.


Both \daf and \ceci adopt the intersection-based method which makes them performs better than edge verification method CFL-Match in most cases. 
Although \fast adopts the edge verification method, it can finish edge verification in one cycle thanks to our pipelining design on FPGA, which makes its cost even less than the intersection-based method on CPU.

Another trend in Fig. \ref{fig_algorithm} is that as data size grows, the acceleration ratio of \fast compared with other three CPU-based algorithms also increases, e.g., for $q3$, the average rate is 26.0x, 33.0x and 59.0x and for $q_8$, the average acceleration rate is 59.0x, 86.0x and 121.0x in DG01, DG03 and DG10, respectively.
It is because the cost of edge verification in \fast remains one cycle while the cost in each \emph{recursive call} grows in other three CPU-based algorithms as the data size grows.

For full comparison, the parallel version \daf-8 and \ceci-8 are also evaluated. However, \daf-8 encounters out of memory error when processing DG03 and DG10. So we only prsent the results of \ceci-8 in Fig. \ref{fig_algorithm}. The average acceleration rate of \fast compared with \ceci-8 is 5.79x, 8.51x and 9.31x in DG01, DG03 and DG10, respectively.

\begin{figure}[htbp]
\centerline{\includegraphics[width=1\columnwidth]{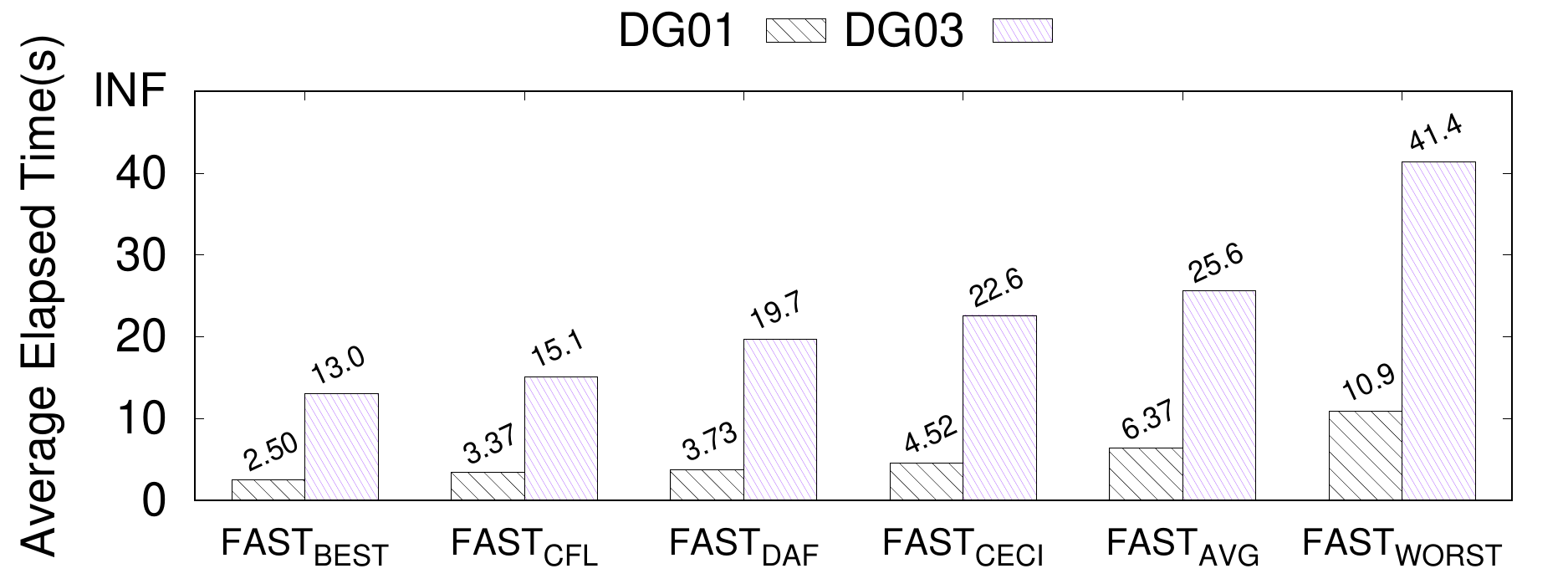}}
\caption{The elapsed time of \fast with different matching orders}
\label{fig_order}
\end{figure}

To evaluate the impact of matching orders, we test \fast with the following orders: (1) \cfl's order; (2) \daf's order; (3) \ceci's order; (4) all other random connected orders. 
The results are illustrated in Fig. \ref{fig_order}.
For each query, we extract the minimum, average and maximum elapsed time denoted as \best, \average and \worst orders, respectively. It can be seen from the figure that the average elapsed time of \fast with \cfl's, \daf's and \ceci's orders is very close to each other. The \fast with \worst matching order can still outperform \cfl, \ceci and \daf (by 9.6x, 11.1x and 36.3x, respectively) which further proves the effectiveness our CPU-FPGA co-designed framework.

\controlspace

\subsection{Scalability Testing}
In this subsection, we evaluate the scalability of our \fast algorithm by using a billion-scale graph DG60. 

\begin{figure}[htbp]
\controlspace
\centerline{\includegraphics[width=1\columnwidth]{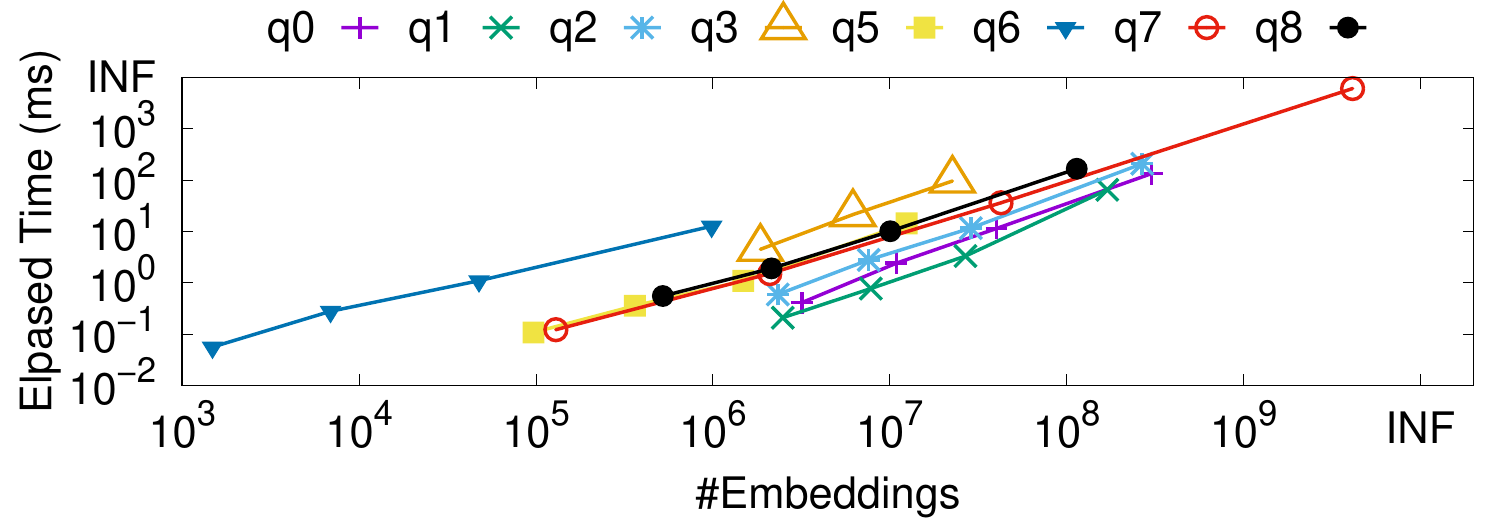}}
\controlspace
\caption{Scalability Testing of \fast (vary $x$)}
\label{fig_scalability}
\controlspace
\end{figure}

\stitle{Varying scale factor}. We run all algorithms on the DG01, DG03, DG10 and DG60. All the other three algorithms fail to finish a single query for the DG60. \emph{CECI} has a segment fault during execution. \emph{CFL-Match} uses an adjacency matrix representation of the data graph to overcome the overhead of edge verification, resulted in out of memory errors for large graphs like DG60. 
As for \emph{DAF}, it encounters overflow errors during execution. 
The problem is caused by the much fewer labels of the \emph{LDBC} datasets (i.e. 11 labels)
which makes the search space larger. \fast completes all queries successfully. The experimental result of \fast is illustrated in Fig. \ref{fig_scalability}. The elapsed time increases linearly with respect to the number of embeddings as the scale factor of $x$ for DG$x$ grows.

\begin{figure}[htbp]
\controlspace
\centerline{\includegraphics[width=1\columnwidth]{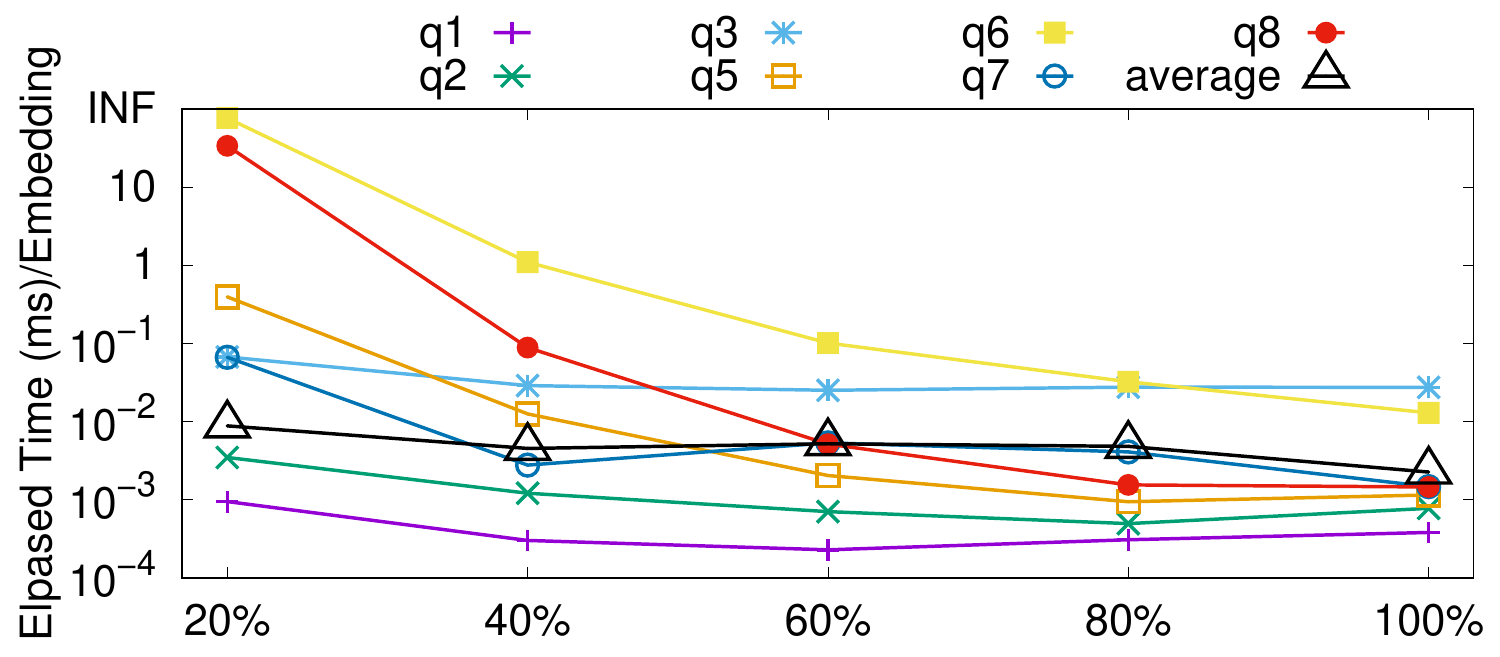}}
\controlspace
\caption{Scalability Testing of \fast (vary $|E(G)|$)}
\label{fig_scalability_edges}
\controlspace
\end{figure}

\stitle{Varying $|E(G)|$}. We keep all vertices and sample 20\%, 40\%, 60\%, and 80\% edges of DG60 uniformly to further test the scalability of \fast. 
Fig. \ref{fig_scalability_edges} indicates that the \textbf{average} elapsed time per embedding has no apparent changing as $|E(G)|$ increases, which verifies the scalability of \fast. The reasons for high elapsed time per embedding for $q5$, $q6$, and $q8$ in the 20\% sample are as follows: (1) The number of embeddings is very small for these queries, e.g., 12 for $q6$ and 36 for $q8$. (2) The cost of data transfer and index construction affects overall performance more apparently, when $E(G)$ is small.

\subsection{Discussion}


\fast algorithm can be easily extended to multi-FPGA environments. Each \cst structure is an independent and complete search space. Combined with our workload estimation method, the CPU can assign the \cst structure to the FPGA with the minimum total workload and collect final results after all the FPGAs complete their tasks. One interesting future work is to combine \fast in the distributed environment to accelerate distributed subgraph matching.

\section{Conclusion} \label{conclusion}
In this paper, we present the first CPU-FPGA co-designed framework to accelerate subgraph matching. Our BRAM-only matching process significantly reduces the costly data transfer between BRAM and DRAM on FPGAs. Moreover, with the workload estimation method of \cst, our framework can be potentially extended to multi-FPGA environment. The experimental results demonstrate that our framework significantly outperforms the state-of-the-art algorithms. In the future, we will investigate integrating \fast into graph database systems and RDF engines to accelerate subgraph queries.

\section*{ACKNOWLEDGMENT}
Xin Jin is supported by 2018YFB1003504. Xuemin Lin is supported by 2018YFB1003504, NSFC61232006, ARC DP200101338, ARC DP180103096 and ARC DP170101628. Shiyu Yang is supported by NSFC61802127 and Shanghai Sailing Program 18YF1406700. Lu Qin is supported by ARC FT200100787.

\bibliographystyle{abbrv}
\bibliography{main.bib}

\end{document}